\documentclass[12pt]{article}

\usepackage{ifpdf}
\usepackage{graphics}
\usepackage{graphicx}
\usepackage{amssymb}
\usepackage{amsmath}
\usepackage{slashed}
\usepackage{cite}
\usepackage{epsfig}
\usepackage{datetime}
\usepackage[usenames, dvipsnames]{color}

\newcommand{\be}{\begin{equation}}
\newcommand{\ee}{\end{equation}}
\newcommand{\bal}{\begin{align}}
\newcommand{\eal}{\end{align}}
\newcommand{\bea}{\begin{eqnarray}}
\newcommand{\eea}{\end{eqnarray}}
\newcommand{\half}{\frac{1}{2}}

\def\p{\partial}

\def\O{\mathcal{O}}

\def\bra#1{\langle #1 |}
\def\ket#1{| #1 \rangle}
\def\vev#1{\langle #1 \rangle}

\def\half{\frac{1}{2}}

\def\p{\partial}
\def\Tr{{\rm Tr}}

\def\RR{{\mathbb R}}

\def\del{\partial}

\def\vk{{\vec k}}

\def\vx{{\vec x}}

\def\CO{{\cal O}}

\def\tr{{\rm tr}}

\def\HIR{{\cal H}_{IR}}
\def\HUV{{\cal H}_{UV}}
\def\PP{\mathbb P}

\def\dt{{\delta t}}
\def\CF{{\cal F}}
\def\CH{{\cal H}}
\def\w{{\omega}}

\newcommand{\bu}{\bar{u}}
\newcommand{\bi}{\bar{i}}
\newcommand{\bj}{\bar{j}}
\begin{document}

\thispagestyle{empty}
\begin{flushright}
BRX-TH 6287\\
\end{flushright}
\vspace*{.05cm}
\begin{center}
{\Large \bf Coarse Grained Quantum Dynamics}

\vspace*{.75cm}{Cesar Agon${}^{a,b}$, Vijay Balasubramanian${}^{c,d,e}$,}\\ 
{Skyler Kasko${}^{f}$, and Albion Lawrence${}^{b}$}\\
\vspace{.5cm} {\em ${}^a$ C.N. Yang Institute for Theoretical Physics,}\\ 
{\em Stony Brook University, Stony Brook, NY, 11794,USA}\\
\vspace{.2cm} {\em ${}^b$ Martin Fisher School of Physics, Brandeis University,}\\ 
{\em Waltham, MA 02454, USA}\\
\vspace{.2cm}
{\em ${}^c$ David Rittenhouse Laboratory, University of Pennsylvania,}\\ 
{\em Philadelphia, PA 19104, USA}\\
\vspace{.2cm}
{\em ${}^d$ CUNY Graduate Center, Initiative for the Theoretical Sciences,}\\ 
{\em New York, NY 10016, USA}\\
\vspace{.2cm}
{\em ${}^e$ Theoretische Natuurkunde, Vrije Universiteit Brussel, and}\\ 
{\em International Solvay Institutes, Pleinlaan 2, B-1050 Brussels, Belgium}\\
\vspace{.2cm}
{\em ${}^f$ Dept. of Physics, University of California,
}\\ 
{\em 
Santa Barbara, CA 93106, USA}

\vspace{1cm}{\bf Abstract}
\end{center}

Inspired by holographic Wilsonian renormalization, we consider coarse graining a quantum system divided between short distance and long distance degrees of freedom, coupled via the Hamiltonian.  Observations using purely long distance observables are described by the reduced density matrix that arises from tracing out the short-distance degrees of freedom.  The dynamics of this density matrix is non-Hamiltonian and nonlocal in time, on the order of some short time scale.  We describe this dynamics in a model system with a simple hierarchy of energy gaps $\Delta E_{UV} > \Delta E_{IR}$, in which the coupling between high-and low-energy degrees of freedom is treated to second order in perturbation theory.  We then describe the equations of motion under suitable time averaging, reflecting the limited time resolution of actual experiments, and find an expansion of the master equation in powers of $\Delta E_{IR}/\Delta E_{UV}$, after the fashion of effective field theory.  The failure of the system to be Hamiltonian or even Markovian appears at higher orders in this ratio.   We compute the evolution of the density matrix in three specific examples: coupled spins,  linearly coupled simple harmonic oscillators, and an interacting scalar QFT.  Finally, we argue that the logarithm of the Feynman-Vernon influence functional is the correct analog of the Wilsonian effective action for this problem.

\eject

\setcounter{tocdepth}{2}
\tableofcontents

\section{Introduction}

Quantum entanglement has emerged as a central concept in the study of the underpinnings of gauge-gravity duality.  The prescription of Ryu and Takayanagi \cite{Ryu:2006bv,Ryu:2006ef}, and its time-dependent generalization \cite{Hubeny:2007xt}, encodes the entanglement entropy between spatial regions in the field theory in the area of minimal or extremal surfaces in the dual spacetime.   Through this, there are good arguments that spatial connectedness in the bulk encodes quantum entanglement of disjoint regions on the boundary \cite{VanRaamsdonk:2010pw,Maldacena:2013xja,Balasubramanian:2014hda}.  

On the other hand, the partitioning of a quantum field theory according to spatial or spacetime scales is fundamental to our physical understanding of quantum field theory, via the renormalization group.   In textbook treatments of renormalization one chooses variables so as to disentangle the ``UV" and ``IR" degrees of freedom.  However, there are many contexts in which one does not do this, or even wish to:
\begin{itemize}
\item As argued in \cite{Balasubramanian:2011wt}, integrating out large Euclidean momenta in a path integral
leads to a reduced density matrix for the IR modes, and the higher-derivative terms are precisely the sign of entanglement between the UV and IR.
\item The entanglement spectrum of a reduced density matrix for low-momentum modes can be a useful way to characterize the long-wavelength behavior of a lattice theory \cite{thomale2010nonlocal}.
\item There is a venerable history of treating ``slow" variables (defined in various ways) as an open quantum system interacting with ``fast modes" to provide a microscopic underpinning of stochastic and hydrodynamic equations. For classic work see \cite{mori1952quantum,mori1953quantum,zwanzig1960ensemble,zwanzig1961memory,zwanzig2001nonequilibrium}.  Some recent work (hardly an exhaustive list!) includes \cite{Jeon:1995zm,Jeon:1994if,Calzetta:1998ng,Meyer:1998dg,Minami:2012hs}.
\item Fluctuations of the cosmic microwave background radiation are analyzed by momentum scale, and different momentum modes are entangled \cite{Lombardo:1995fg,Calzetta:1999zr,Burgess:2006jn,Akkelin:2013jsa,Burgess:2014eoa,Burgess:2015ajz, Shandera:2017qkg}.
\item It is useful to treat the IR region of jets in high-energy particle collisions as an open quantum system -- {\it cf. } \cite{Neill:2015nya} and the references therein.
\end{itemize}

One important setting where choosing variables to ``disentangle" the UV and IR degrees of freedom can obscure the physics is  in the AdS/CFT correspndence.  In this case, 
there is ample evidence that spacetime scale in a QFT is related via gauge-gravity duality to the radial direction in the dual asymptotically anti-de Sitter space \cite{Maldacena:1997re,Susskind:1998dq}, with the region of anti-de Sitter space close to the boundary dual to the UV region of the quantum field theory (i.e. ``scale-radius duality").  Various prescriptions have emerged for relating the radial evolution of bulk fields to the renormalization group flow of the dual field theory \cite{Akhmedov:1998vf,Balasubramanian:1999jd,deBoer:1999xf,Porrati:1999ew,Klebanov:2000hb,Heemskerk:2010hk,Faulkner:2010jy}. Of these, the Wilsonian prescription of \cite{Heemskerk:2010hk,Faulkner:2010jy}\ lends itself most readily to a finite-N generalization \cite{Balasubramanian:2012hb}.  In this scheme, the IR and UV regions are clearly entangled \cite{Balasubramanian:2012hb}.  

In this paper we will explore such open quantum systems from the quantum mechanical/quantum field theoretic point of view, with the eventual aim of shedding light on scale-radius duality.  Before doing this, let us recall the discussion in \cite{Balasubramanian:2012hb}.

The AdS/CFT correspondence states that the dual of a $d$-dimensional large-N conformal field theory (N could be the rank of a gauge group, or the central charge of a 2D CFT) is string- or M-theory in $AdS_{d+1} \times X$, where $X$ is some space with constant positive curvature.  For CFTs on $\RR^{1,d-1}$, one considers a Poincar\'e patch of anti-de Sitter space, with coordinates
\be
	ds^2 = R^2 \frac{dr^2}{r^2} + \frac{r^2}{R^2} dx_d^2
\ee
Here $dx_d^2$ is the flat metric on $d$-dimensional Minkowski space; $R$ is the radius of curvature of $AdS_d$.
To implement a renormalization group flow after the fashion of Wilson, 
Refs. \cite{Heemskerk:2010hk,Faulkner:2010jy}\ propose the following.  The cutoff $\Lambda$ is associated with a definite radial coordinate, $r_{\Lambda}  = R^2 \Lambda$. One breaks up the path integral over fields propagating on $AdS_d$ into modes with $r > r_{\Lambda}$ and $r < r_{\Lambda}$; interprets the path integral for $r < r_{\Lambda}$ with fixed fields at $r = r_{\Lambda}$ as the generating functions of correlators in the cutoff theory; and integrates this over the field values at $r = r_{\Lambda}$ weighted by the path integral over the fields for $r > r_{\Lambda}$.  

In this procedure, nontrivial operators are induced at the cutoff {\it even when the theory is an unperturbed conformal field theory} \cite{Balasubramanian:2012hb}.  In particular, at a given cutoff $\Lambda$, one induces terms in the Wilsonian action of the form
\be
	\Delta S_{\Lambda} = \int d^d x d^d y \gamma_{ab}(x - y; \Lambda) \O_a(x) \O_b(y) \label{eq:nonlocalterm}
\ee
where $\O_a$ correspond to single-trace operators dual to supergravity fields.  The kernel $\gamma$ is nonlocal over spacetime distances of order $\Lambda^{-1}$.  In the holographic picture, these operators have a clear interpretation \cite{Balasubramanian:2012hb}.  If one excites the bulk in the ``infrared" region $r < r_{\Lambda}$, these excitations can propagate out to the region $r > r_{\Lambda}$.  The induced term (\ref{eq:nonlocalterm}) acts precisely to describe the transfer of these modes between the IR and UV regions on time scales of order $\Lambda^{-1}$. In other words, they take care of the fact that the IR region comprises an open quantum system.\footnote{Note that the relationship between this holographic cutoff and any factorization of the Hilbert space is an open question (see for example \cite{Balasubramanian:2013lsa}\ for a discussion).  In large-N vector models dual to higher-spin theories in anti-de Sitter space, the associated cutoff appears to be a point-splitting cutoff on gauge-invariant bilocal operators \cite{Mintun:2014gua}.  This is an important issue that we will put aside for the present.} 

In this work we will study simple quantum systems which capture the spirit of the split between infrared and ultraviolet modes seen in quantum field theories. We will focus on theories with a hierarchical structure of energy levels governed by level splittings $\Delta E_{IR} \ll \Delta E_{UV}$. This structure, the underpinning of the Born-Oppenheimer approximation, is the basis of Wilson's pioneering work \cite{Wilson:1965zzb,Wilson:1974mb,Wilson:1993dy,Glazek:1993rc}, and provides a conceptual underpinning for effective field theory \cite{Moody:1989vh}.

 As we will argue, experiments with limited spatial resolution are described by an ``IR density matrix", a reduced density matrix which arises from tracing out short-distance modes.  However, realistic experiments also have limited resolution in time, so we will implement a straightforward, physical time-averaging procedure to describe them. We will compute the master equation describing the time evolution of the time-averaged IR density matrix. We will find that the master equation can be organized in a power series in $(\Delta E_{IR}/\Delta E_{UV})$ after the fashion of effective field theory, for which we can begin to identify parallels with the discussion in \cite{Balasubramanian:2012hb}.  

The study of open quantum systems is a well-developed subject  (see the reviews \cite{Leggett:1987zz,Calzetta:1999zr,breuer2007theory}, e.g., the treatment of fast or ultraviolet modes as an environment for slow, infrared modes.    Our work contributes an abstract treatment  that leads to an effective field theory-like expansion of the master equation for reduced density matrices of subsystems.\footnote{A notable exception is the recent work \cite{2014AnPhy.345..141D,2014arXiv1405.2077D}, which treats the ``IR" mode classically, and derives a dissipative dynamics for that mode. Another is the related set of papers which consider the density matrix for low-mass fields after integrating out high-mass fields \cite{Calzetta:1996sy,Boyanovsky:2015tba,Boyanovsky:2015xoa}. These are complementary to the present work.} We focus on this abstract language for two reasons.  First, it highlights essential physics -- the presence of a hierarchy of energy scales.  Secondly, an abstract approach is best suited to our goal of understanding gauge-gravity duality, for which the variables that appear in a path integral approach to the gauge theory have by themselves no clear dual (to begin with they are not even gauge invariant). 

In the following section, we will embark on our computation of the master equation in perturbation theory for a simple quantum system motivated by the essential structure of quantum field theories. After implementing a physical time-averaging, we will see a Born-Oppenheimer-type expansion emerge, in which non-Hamiltonian and non-Markovian dynamics appear starting at second order in $1/\Delta E_{UV}$. We provide a simple expression for the time evolution of R\'enyi entropies of the IR density matrix.  We will work two simple examples, a coupled spin system and the Caldeira-Leggett model \cite{Caldeira:1982iu}\ at zero temperature, to see how this expansion plays out, and then describe the master equation for scalar quantum field theories with cubic interactions.  Such master equations for low-spatial-momentum modes have been computed for four-dimensional theories via the influence functional approach \cite{Lombardo:1995fg}, and our work further contributes a Born-Oppenheimer-like framework and a time-averaging procedure for understanding the effects of finite time-resolution.

In the conclusions we will draw what lessons we can for gauge-gravity duality, relating the appearance of non-Hamiltonian and non-Markovian terms to nonlocal terms that emerge \cite{Balasubramanian:2012hb}\ in the Wilsonian approach of \cite{Heemskerk:2010hk,Faulkner:2010jy}.  We will argue that the natural framework for understanding this work would be to use the results of \cite{Heemskerk:2010hk,Faulkner:2010jy}\ to compute an influence functional \cite{Feynman:1963fq,feynman2012quantum}\ for the IR modes, and note further that the logarithm of the influence functional is the natural extension to the Wilsonian effective action to understanding finite-time processes. We will then provide some further speculations regarding the use of these results for understanding gauge-gravity duality.

In the appendices we review some concepts which may be unfamiliar for some of our audience (while being bread and butter to others). First, we address a common confusion we encountered when discussing this work, that textbook treatments of renormalization do not consider entanglement between UV and IR modes. We review various approaches to renormalization of quantum field theories to explain how, in those cases, these modes are disentangled. Next, we discuss some issues with states with initial entanglement. We then discuss a an obstruction to computing the von Neumann entropy for the IR density matrix, in perturbation theory. Finally, we review the path integral approach to computing the dynamics of density matrices, and in particular the Feynman-Vernon influence functional.

\subsection{Update from previous versions}

The present version of this paper is a substantial rewriting of an earlier draft which appeared in December 2014. The essential calculations and physical conclusions have not changed.  We have reorganized the paper to make our motivations and results clearer, and added one new example which is explored in greater depth in a follow-up paper \cite{Agon:2017oia}.  Since the first version of this work appeared on the arxiv, a number of interesting papers on related subjects have appeared, including \cite{Boyanovsky:2015xoa,Boyanovsky:2015tba,Neill:2015nya,Boyanovsky:2015jen,Burgess:2015ajz,Shandera:2017qkg,Burgess:2017ytm,Boyanovsky:2018fxl}.

%\section{Varieties of Wilsonian renormalization}

\section{Dynamics of $\rho_{IR}$ in perturbation theory \label{2}}

\subsection{Motivation}

Consider an interacting scalar field theory
\be
	H = \frac{1}{2} \pi_{\phi}^2 + \frac{1}{2} (\del\phi)^2 + \frac{1}{2} m^2 \phi^2 + \frac{\lambda}{4} \phi^4\label{eq:phifourh}
\ee
defined on some lattice with spacing $a$.  Now consider  a measuring device which directly couples to $\phi$, but has finite resolution in space and time. That is, if we write the field $\phi$ in the Schr\"odinger picture as
\be
	\phi(x) = \sum_{k = - \pi/a}^{\pi/a} \frac{1}{2\sqrt{\pi \omega(k)}} a_k e^{ik\cdot x} + {\rm h.c}\ ,
\ee
then our measuring devices couple to $a_k$, $a_k^{\dagger}$ for $|k| < \Lambda \ll \frac{\pi}{a}$, and record the time of the measurement with temporal accuracy $\delta t$.

The Hilbert space can be broken up into
\be
	{\cal H} = {\cal H}_{IR} \otimes {\cal H}_{UV}\label{eq:hsdecomp}
\ee
where ${\cal H}_{IR}$ is generated by $a^{\dagger}_{k}$ for $|k| < \Lambda$ and ${\cal H}_{UV}$ is generated by $a^{\dagger}_k$ for $|k| > \Lambda$. Note that we are not directly breaking up the Hilbert space according to energy scale.  Firstly, for an interacting theory, spatial momentum and energy will not be directly related.  Secondly, we may be interested in high-energy objects made up of many low-energy quanta.  After all, the physics of the sun is well described by the standard model cutoff at a TeV, even though its total mass is of order $10^{54}\ GeV$.\footnote{One may, however, wish to restrict the Hilbert space to states with low energy {\it density} of order $\Lambda^D$, where  $D$ is the space-time dimension. }

We imagine an experiment of the following form.  Begin with the system in its exact ground state $\ket{0}$, and act on it with some infrared operator ${\cal O}_{IR}$.  Let the resulting state evolve in time,
\be
	\ket{\psi(t)} = e^{- \frac{i}{\hbar} H t} {\cal O}_{IR} \ket{0}\ .\label{eq:pgsinitial}
\ee
Now compute the probability of measuring the IR degrees of freedom in some state $\ket{a} \in {\cal H}_{IR}$.  We are not making any measurements in ${\cal H}_{UV}$, so we should sum the probabilities over all possible final states in ${\cal H}_{UV}$.  The result is
\begin{eqnarray}
	P(a,t) & = & \sum_{\ket{u} \in \HUV} \big| \bra{u}\bra{a} e^{- \frac{i}{\hbar} H t} {\cal O}_{IR} \ket{0} \big|^2\nonumber\\
	& = & \tr \PP_a e^{- \frac{i}{\hbar} H t} {\cal O}_{IR} \ket{0} \bra{0} {\cal O}_{IR}^{\dagger} e^{\frac{i}{\hbar} H t}\nonumber\\
	& = & \tr_{\HIR} \PP_a \rho_{IR}(t) \label{eqn:measurement}
\end{eqnarray}
where $\PP_a = \ket{a}\bra{a}$, and
\be
	\rho_{IR}(t) = \tr_{\HUV} \left[ e^{- \frac{i}{\hbar} H t} \ket{\psi(0)}\bra{\psi(0)} e^{\frac{i}{\hbar} H t}\right]\label{eq:rdm}
\ee
More generally, the expectation value at time $t$ of measurements of $A_{IR}$ acting on $\HIR$ is $\vev{A} = \tr A \rho_{IR}(t)$. Based on this we take $\rho_{IR}$ to be the object of interest.  

\subsection{Setup}

Wilson emphasized \cite{Wilson:1974mb,Wilson:1974mb,Glazek:1993rc}\ that the energy spectrum for quantum field theories has a hierarchical structure, as illustrated in Figure 1.  In order to focus on the effects of this structure, we work with a simpler abstract model that captures it. That is, we consider a Hilbert space with product structure
\be
	{\cal H} = {\cal H}_{IR} \otimes {\cal H}_{UV}
\ee
and a Hamiltonian of the form
\be
	H = H_{IR}\otimes {\bf 1} + {\bf 1}\otimes H_{UV} + \lambda V\label{eq:splithamiltonian}
\ee
where $V$ acts on both ${\cal H}_{IR}$ and ${\cal H}_{UV}$.

\begin{figure}
\centering
\includegraphics[scale=.35]{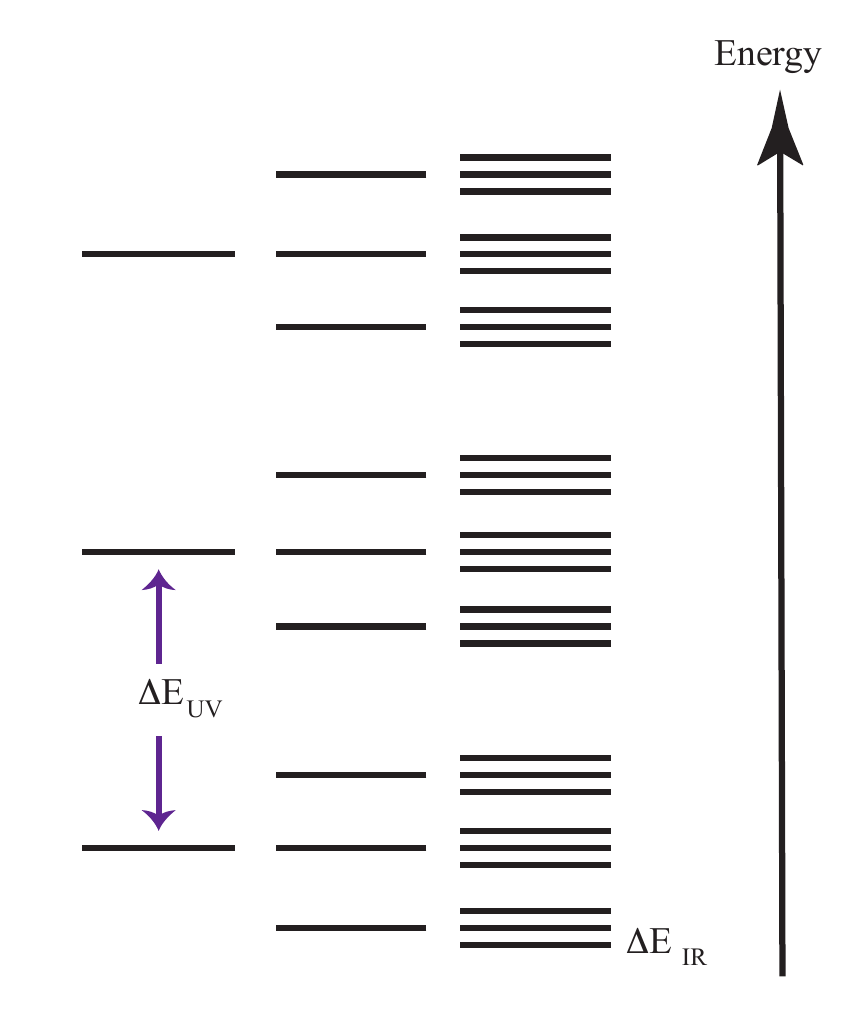}
\caption{Cartoon of band structure: large jumps correspond to UV quanta, small jumps to IR quanta.}
\end{figure}

Our goal is to compute the ``master equation" for $\rho_{IR}(t)$, that is, the right hand side of the expression 
\be
	i \hbar \p_t \rho_{IR}= {\cal L}(\rho)\,.
\ee
where, ${\cal L}(\rho)\,$ represents a differential operator on $\rho_{IR}$, which depends on the full $\rho$ through the intial conditions.  Recognizing that measurements in the IR theory will typically have limited time resolution, we will also determine the time-evolution of a version of $\rho$ that is coarse-grained  by time averaging. 

We will assume that:
\begin{enumerate}
\item {\it The interaction $\lambda V$ can be treated perturbatively.}   Specifically we will work to order $\lambda^2$.  In perturbation theory starting from Fock space there is a tight connection between momentum and energy.  Note that perturbation theory for the density matrix itself can fail at long times, due to secular terms in the perturbative expansion \cite{fleming2011accuracy}.
\item {\it The eigenvalue spacing for $H_{UV}$, $H_{IR}$ can be characterized by scales $\Delta E_{UV} \gg \Delta E_{IR}$.}   This is a simplification.  In general, we expect a local system to have a nested hierarchy of energy levels, as shown in Figure 1, corresponding to different momentum modes of the fundamental fields. That being said, at this order of perturbation theory, we will find that we can also compute the master equation for cubic scalar quantum field theories.
\item {\it Factorized initial states}.  Following much of the literature on open quantum systems, we will consider initial states for which the UV and IR degrees of freedom are not entangled, so that $\rho_{IR}(0)$ is a pure state. These have a master equation which is local in time.  This is of course not the most general situation -- small excitations of the ground state such as (\ref{eq:pgsinitial})\ will in general be highly entangled between the UV and IR -- but it can arise in interesting physical situations.  For example, if we prepare the state by measuring the IR with a non-degenerate Hermitian operator, the state will collapse to a product state.  Another well-studied situation is the ``interaction quench", in which $\lambda$ is suddenly turned on at $t = 0$. The entanglement of spatial regions after a quench has been well-studied, beginning with the pioneering work of \cite{Calabrese:2005in,Calabrese:2006rx,Calabrese:2007rg}.  
\end{enumerate}

%The reader may feel that this is far from our original motivations based on gauge-gravity duality: we are working with a perturbative system, and we have stepped back somewhat from studying manifestly local quantum field theory.  We have kept what we feel is an essential feature of such theories, and we work with perturbation theory because we can do calculations readily and still learn something about the dynamics of entanglement across scales.  In the long run, a more abstract approach will be most useful in studying gauge-gravity duality.

\subsection{Perturbative calculation}

The calculation of the master equation for $\rho_{IR}(t)$ can be done by, e.g., projection operator techniques (see \cite{breuer2007theory}).    To be self-contained, we will re-derive  results from the theory of open quantum systems in manner consistent with our approximations and perspective.
%We expect many of our intended readers will not be familiar with the literature on open quantum systems, so we will carry out a particular derivation here consistent with our approximations and perspective. 
%In our truncated model, the entanglement entropy oscillates as excitations mix between the IR and UV degrees of freedom; however, the discussion in \cite{Balasubramanian:2012hb}\ suggests that there could be smooth evolution, local in some variable characterizing spatial scales.

We consider $\ket{\Psi(0)} = \ket{\psi_{IR}}\ket{\bu}$, where $u$ labels eigenstates of $H_{UV}$; $\bu$ is some particular state, possibly but not necessarily the ground state, while $\ket{\psi_{IR}}$ is taken to be some arbitrary state in $\HIR$. In general, when the initial state of a coupled system and environment is factorized between the two, the reduced density matrix of the system satisfies a master equation which is local in time (see \cite{breuer2007theory,BreuerFoundations:2012,2014RPPh...77i4001R}\ for discussion and references):
\begin{eqnarray}
	i \hbar \p_t \rho_{IR}(t) & = & [H_{eff}(t), \rho_{IR}(t)] + i \left\{A(t), \rho_{IR}(t)\right\} + \gamma[\rho_{IR}(t)]\nonumber\\
	& \equiv & [H_{eff}, \rho_{IR}(t)] + \Gamma[\rho_{IR}(t)] \label{eq:tlocevol}
\end{eqnarray}
$\Gamma$ labels the non-Hamiltonian part of the master equation for $\rho_{IR}$, with
%  The components of $\Gamma$ can be written as:
%  
\begin{eqnarray}
	A(t) & = & - \half \sum_{k} h_{kl} (t) L^{\dagger}_l(t) L_k(t)\nonumber\\
	\gamma[\rho_{IR}] & = & i \sum_k h_{kl}(t) L_k(t) \rho_{IR}(t) L^{\dagger}_l(t)\label{eq:tlmaster}
\end{eqnarray}
where $L_k$ are some set of operators that can depend on the initial state of the UV degrees of freedom but act on the IR, and $h_{kl}$ is a Hermitian matrix.\footnote{Such state dependence is not usually discussed in Wilsonian renormalization: in particle physics examples, one is usually assuming that the UV theory is in the ground state.  More generally, the Born-Oppenheimer discussion in Appendix A shows that even the effective Hamiltonian (\ref{eq:boheff})\ depends on the state of the UV modes.}   Here $k,l$ are arbitrary indices that index the $L_k$; they do not have to have any particular relation to the UV or IR Hilbert spaces.  This is almost the Kossokowski-Lindblad equation for Markovian dynamics\cite{Kossakowski1972247,Lindblad:1975ef}.  However, Markovian dynamics requires that the eigenvalues of $h$ be positive, and this condition is well known to fail in general.\footnote{In the Kossakowski-Lindblad equation, following from the assumption that the time-evolution of $\rho$ is described by a completely positive dynamical semigroup, $h, L$ are time-independent. However, a more general definition of Markovian includes divisible dynamical maps \cite{2010PhRvA..81f2115L}, in which $h_k$, $L_k$ can be time-dependent, but the eigenvalues of $h$ remain positive.}   Thus (\ref{eq:tlocevol}) can be non-Markovian even if it looks local in time, because there can be history dependence hidden in the operators $A$ and $\gamma$ diagnosed by the breakdown of positive definiteness of $h_{ij}$  \cite{breuer2007theory,BreuerFoundations:2012,2014RPPh...77i4001R}. 

We will construct (\ref{eq:tlocevol}) to second order in perturbation theory, using the fact that the finite-time evolution of $\rho_{IR}$, simply denoted as $\rho$ hereafter, has a Kraus representation (see for example \cite{BreuerFoundations:2012,2014RPPh...77i4001R}).  Thus
\be
	\rho_{IR}(t) \equiv \rho(t) = \sum_{\alpha} K_{\alpha}(t) \rho(0) K_{\alpha}^{\dagger}(t)\label{eq:evfac}
\ee
in terms of certain operators $K_\alpha$ that can be derived from the time evolution.\footnote{The Kraus representation guarantees that the map $\rho(0) \to \rho(t)$ is ``completely positive". This representation is possible when the initial state is disentangled between the IR and the UV.  For intermediate times $t'$, the state will be entangled, and the map from $\rho(t') \to \rho(t)$ will not be completely positive. This may include $t'$ arbitrarily close to $t$, as diagnosed by the non-positivity of the eigenvalues of $h$; this non-positivity for infinitesimal time evolution means that the evolution will not be Markovian, as entanglement has been generated.}
In our example, the density matrix $\sigma(t) \equiv \ket{\Psi(t)}\bra{\Psi(t)}$ in the full Hilbert space $\HIR\times\HUV$ satisfies unitary evolution:
\be
	\sigma(t) = U(t) \sigma(0) U^{\dagger}(t)\ ,
\ee
where 
\be
	U(t) = e^{- i (H_{IR} + H_{UV})t } T e^{- i \lambda \int_0^t dt' V_I(t')}\ ,
\ee
$V_I$ is the perturbation in the interaction picture, and $T$ is the time-ordering operator.  The time evolution of $\rho(t)$ in our case is:
\begin{eqnarray}
	\rho(t)&=&\textrm{Tr}_{UV}\sigma(t)=\textrm{Tr}_{UV}U(t,0)\sigma(0)U^{\dagger}(t,0)\nonumber \\
	&=&\sum_{u}\langle u|U(t,0)|\bu\rangle |IR\rangle\langle IR|\langle \bu | U^{\dagger}(t,0)|u\rangle \nonumber \\
	&=&\sum_{u}\langle u|U(t,0)|\bu\rangle \rho(0)\langle \bu | U^{\dagger}(t,0)|u\rangle 
\end{eqnarray} 
where $| \bar{u}\rangle $ and $|IR\rangle$ are the UV and IR parts of the factorized initial state, $|u\rangle$ is a basis for the UV Hilbert space with one basis element being $|\bar{u}\rangle$, and $\rho(0) = |IR\rangle\langle IR |$ is the initial reduced density matrix for the IR modes.   So the Kraus operators can be taken to be
\be
 K_{u}=\langle u|U(t,0)|\bu\rangle\
\ee
with $u$ indexing UV degrees of freedom, and $|\bar{u}\rangle$ being the UV part of the initial state.  Thus the Kraus operators are treated as  functions of $u$ that  are {\it initial-state dependent}, but we have suppressed the state-dependence in our notation.

We can rewrite  the master equation (\ref{eq:tlocevol},\ref{eq:tlmaster}) in terms of the Kraus operators.  First,  we can perturbatively expand
\begin{eqnarray}
 	\rho(t) & = & \rho^{(0)}(t) + \lambda \rho^{(1)}(t) + \lambda^2 \rho^{(2)}(t) + \ldots\nonumber\\
	H_{eff} & = & H_{IR} + \lambda H_{eff}^{(1)} + \lambda^2 H_{eff}^{(2)} + \ldots \nonumber\\
	A & = & \lambda A^{(1)} + \lambda^2 A^{(2)} + \ldots \nonumber\\
	\gamma & = & \lambda \gamma^{(1)} + \lambda^2 \gamma^{(2)} + \ldots
\end{eqnarray}
Here $\rho^{(0)}$ is the density matrix of the initial IR state evolved in time by the IR Hamiltonian $H_{IR}$.
The master equation for time evolution can be determined by representing $\rho(t)$ in the Kraus representation and expanding $K_u$ in the perturbation.
% We then compute $H_{eff}, A, \gamma$ by expanding (\ref{eq:tlocevol}) in a power series in $\lambda$ and equating the left- and right-hand sides. 
We find to ${\cal O}(\lambda^2)$:
\begin{eqnarray}
	K_{\bu} & = & \exp \left\{ - i \left(H_{eff} + i A\right)t\right\}\nonumber\\
	\gamma & = & i \p_t \sum_{u\neq \bu} K_u(t) \rho(0) K_u^{\dagger}(t)
\end{eqnarray}
where $K_{\bar{u}}$ is a partial matrix element for transitions between an initial UV state $\bar{u}$ to itself;  $\gamma$ controls transitions out of $\bar{u}$ into other UV states, while $A$ describes the associated loss of unitarity in the subspace $\ket{\bu}\otimes\HIR$.   The sum in expression for $\gamma$ runs over the part of the UV Hilbert space that is orthogonal to the initial state $|\bar{u} \rangle$.

Since $K_{u\neq \bu}$ is nonvanishing only at ${\cal O}(\lambda)$ and higher, $\gamma = \lambda^2 \gamma^{(2)} + \ldots$.  The form of the time-local master equation (and direct computation) also shows that $A = \lambda^2 A^{(2)} + \ldots$.  Thus, to order $\CO(\lambda)$ we get simply a correction to the effective Hamiltonian in the master equation:
\be
	H^{(1)}_{eff} = \bra{\bu} V \ket{\bu}\label{eq:firstorderham}
\ee
At order $\CO(\lambda^2)$, $A^{(2)},\gamma^{(2)}$ can be written in the form (\ref{eq:tlmaster}) by choosing the indices $k,l$ to each run over the composite index $k,l = um$ with the index $u\neq \bar{u}$ running over a basis for the part of the UV Hilbert space which is orthogonal to the initial state, and $m \in \{1,2\}$.   With this notation, we can write the operators on the right hand side of (\ref{eq:tlmaster}) that define $A$ and $\gamma$ as
\begin{eqnarray}
	L_{u1} & = &  \bra{u} V \ket{\bu}\\
	L_{u2} & = & \int_0^t dt' \bra{u} V_I(t'-t) \ket{\bu}\label{eq:lindbladish} \ ,
\end{eqnarray}
where $V$ and $V_I$ are the interaction in the Schrodinger and interaction pictures respectively.  The effective Hamiltonian and other operators that govern the master equation at this order can be written in terms of $L_{um}$:
\bea
	H^{(2)}_{eff} 
	&=& - \frac{i}{2} \sum_{u\neq {\bar u}} h_{u1,u2} \left(L_{u1}^{\dagger} L_{u2} - L_{u2}^{\dagger} L_{u1}\right)\\
	A^{(2)} 
	&=& - \frac{1}{2} \sum_{u\neq {\bar u}} h_{u1,u2} \left(L_{u1}^{\dagger} L_{u2} + L_{u2}^{\dagger} L_{u1}\right)\\
	\gamma^{(2)} 
	&=& i \sum_{u\neq {\bar u}} h_{u1,u2} \left(L_{u1} \, \rho^{(0)} \,  L_{u2}^{\dagger}  + L_{u2} \, \rho^{(0)} \, s L^{\dagger}_{u1}\right)\label{eq:tlham}
\eea
where $h_{um,u'm'} = \delta(u,u') |m-m'|$ and $\rho^{(0)}(t)$ is the density matrix of the initial IR state evolved in time by the IR Hamiltonian $H_{IR}$.
%We take $k,l$ to run over $(u \neq {\bar u}, a \in \{1,2\})$, set $h_{u1,u2} = h_{u2,u1} = 1/\tau$, and define
%
\vskip .2cm
\noindent{\it Expression in terms of UV correlation functions:}   
There is an elegant expression for the various terms in the master equation in terms of correlation functions of UV operators.
%These quantities can be conveniently expressed in terms of correlation functions of UV operators.  If we write 
Let us write
\be
	\lambda V = \sum_a \Phi_a {\cal O}_a\ ,
\ee
 where ${\cal O}_a, \Phi_a$ are sets of UV and IR operators respectively.     Using the path integral formalism in Appendix D, and defining the connected 2-point function
\be
	G^W_{ab}(t',t'') = \langle {\cal O}^I_a(t'){\cal O}^I_b(t'')\rangle-\langle {\cal O}^I_a(t')\rangle \langle {\cal O}^I_b(t'')\rangle\,,
\ee
where $\langle {\cal O}^I (\tau) \rangle =\Tr_{UV}{\cal O}^I(\tau)\rho_{UV}(0)$  with $\rho_{UV}(0) = |\bar{u}\rangle\langle \bar{u}|$,\footnote{The super index $I$ refers to the fact that those operators are described in the interaction picture of quantum mechanics}
we find that (\ref{eq:tlmaster}) can be written in terms of the following set of operators\footnote{This framework generalizes easily to include an arbitrary initial density matrix for the UV degrees of freedom, again assuming the density matrix for the full system is factorized between UV and IR at $t = 0$.}  
\begin{eqnarray}
	h_{am;bm'} & = & \lambda^2 \, \delta_{ab} \, |m-m'|
	%	\left(\begin{array}{ll} 0 & \frac{1}{\tau^*} \\ \frac{1}{\tau} & 0 \end{array}\right)
	\nonumber\\
	L_{a1} & = &  \Phi_a(0) \nonumber\\
	L_{a2} & = & \int_0^t d\tau \, G^W_{ab}(t,\tau) \, \Phi_b(\tau-t)\ ,
\end{eqnarray}
and
\begin{eqnarray}
	H^{(2)} & = & i \sum_{ab} \int^t_0 d\tau \left( G^W_{ab}(\tau,t) \, \Phi^I_a(\tau-t) \, \Phi^I_b(0) - G^W_{ab}(t,\tau) \, \Phi^I_a(0) \, \Phi^I_b(\tau-t)\right)\nonumber\\
	A^{(2)} & = & \sum_{ab} \int^t_0 d\tau \left( G^W_{ab}(\tau,t)  \, \Phi^I_a(\tau-t) \, \Phi^I_b(0) + G^W_{ab}(t,\tau) \, \Phi^I_a(0) \, \Phi^I_b(\tau-t)\right)\nonumber\\
	\gamma^{(2)} & = & i \sum_{ab} \int^t_0 d\tau \left( G^W_{ab}(\tau,t) \, \Phi^I_b(0)  \, \rho^{(0)}(t) \, \Phi^I_a(\tau-t) 
	\right. \nonumber\\
	& & \ \ \ \ \ \ \ \ \ \ \left. + 
		G^W_{ab}(t,\tau) \Phi^I_b(\tau-t) \, \rho^{(0)}(t) \, \Phi^I_a(0)\right)\ .\label{eq:notimeav}
\end{eqnarray}

\vskip .2cm
\noindent{\it Non-Markovianity}.  A natural question is whether the evolution, packaged in this form, is Markovian.  It is well known to those who study coarse-grained quantum systems that it is not, in general.  Assume the particularly simple case that $V$ commutes with $H_{IR}$ but not with $H_{UV}$ (when $[V,H_{UV}] = 0$, $\Gamma$ vanishes).  A short calculation 
of the non-Hamiltonian part of the master equation gives
\begin{eqnarray}
	& & \sum_{u\neq \bu; m; m'} h_{um,um'} \left(L_{um} \rho(t) L_{um'}^{\dagger} - \half \left\{ L^{\dagger}_{um} L_{um'}, \rho\right\} \right) \nonumber\\
	& & \ \ \ \ \	= \sum_{u\neq \bu} \frac{2 \sin (E_{u\bu} t)}{E_{u\bu}} \left(\bra{u}V\ket{\bu} \rho(t) \bra{\bu}V^{\dagger}\ket{u} - \half \left\{\bra{\bu}V^{\dagger}\ket{u}\bra{u}V\ket{\bu},\rho(t) \right\}\right)\nonumber\\
\end{eqnarray}
where $E_{u\bu} = E_u - E_{\bu}$, and $E_u$ is the energy of $\ket{u}$ with respect to $H_{UV}$. In general, if $E_{u\bu} \geq \Delta E_{UV}$ for $u\neq \bu$, the sine term will lead to oscillations at the scale $\Delta E_{UV}$.  Since the matrix elements in the sum are therefore not positive definite, we know on general grounds \cite{BreuerFoundations:2012,2014RPPh...77i4001R} that the evolution is not Markovian.

\vskip .2cm
\noindent{\it Evolution of UV-IR entanglement}. One of the motivations for this work was to understand the structure of entanglement between the UV and IR. It is straightforward to see that entanglement evolves precisely because of the non-Hamiltonian part of the evolution.  For reasons we discuss in Appendix C (see also \cite{Balasubramanian:2011wt}), the von Neumann entropies are difficult to compute in perturbation theory, but the R\'enyi entropies
\be
	S_{n}(t)=-\frac{\ln \textrm{Tr}\rho^n(t)}{n-1}\,.
\ee
are easily seen to satisfy the equation
\be
	\frac{dS_n(t)}{dt}=\frac{in}{n-1}\frac{\textrm{Tr}[\rho^{n-1}(t)\Gamma(t)]}{\textrm{Tr}\rho^n(t)}\ \label{eq:secondorderrenyi}
\ee
%

%Further insight into when the time evolution of $\rho$ is Markovian comes by rewriting $H^{(2)},A^{(2)}$ and $\gamma^{(2)}$ in terms of correlation functions of the UV parts of $V$.  This representation emerges naturally in the path integral formalism, so we will discuss it further in \S4.  

\subsection{Time averaging}

%\subsubsection{Time averaging operator products}

Realistic apparati have limited accuracy in specifying the time that a given measurement takes place.  To find the probability of a given outcome, one should average $P(a,t)$, the probability of outcome $a$ at a time $t$, over a time interval determined by an appropriate window function $f_{\delta t}(\tau, t)$ where $t$ is the peak of the window function and $\delta t$ is the width.  A typical example is a Gaussian
\be
	f_{g,\delta t}(\tau - t) = \frac{1}{\sqrt{\pi}\delta t} e^{- (\tau-t)^2/\dt^2}\ .
\ee
With this normalization, the sum of (\ref{eqn:measurement}) over all possible orthogonal outcomes ($a$) is equal to 1.
Given a time-dependent function $F(t)$, we denote the time average as:
\be
\overline{F(t)}=\int d\tau f_{\dt}(\tau,t) F(\tau)
\ee
In the case of the Gaussian window function, this expression can be written in Fourier space as:
\be
	\overline{F(t)}=\int \frac{d\omega}{\sqrt{2 \pi}}e^{-\omega
^2\dt^2/4}e^{i\omega t} \tilde{F}(\omega)
\ee
As expected, there is a sharp exponential cutoff for $\omega > \dt^{-1}$.

In applying this averaging to (\ref{eq:tlocevol}), we will consider $\Delta E_{IR} \ll \dt^{-1} \equiv E_c \ll \Delta E_{UV}$.  Thus we will throw away terms in (\ref{eq:tlocevol}) which have
frequencies of $O(\Delta E_{UV})$ as these will be exponentially suppressed after time averaging.  We will, however, keep terms of order ${\cal O}\left[\left(\Delta E_{IR}/E_c\right)^k\right]$.

We now wish to compute a master equation for the time-averaged density matrix ${\bar \rho}(t)$
The nontrivial time dependence of the terms in (\ref{eq:tlocevol}) arises from $L_{u,2}$ in (\ref{eq:lindbladish}).  If we study a matrix element of $L_2$ in the basis $\ket{i}$ of IR eigenstates with energies $E_i$, we find that
\be
	\bra{i} L_{u,2} \ket{j} = \frac{1 - e^{- i (E_{u\bu} + E_{ij})t}}{i(E_{u\bu} + E_{ij})}\bra{i}\bra{u} V \ket{\bu}\ket{j}
\ee
where $E_{ij} = E_i - E_j$. The first term will, in general, survive time averaging.

We find by construction that to second order in perturbation theory in $V$, the time-averaged evolution equation \footnote{ Up to second order in perturbation theory we find that equation (\ref{eq:timeaveqn}) is valid, namely, the time average of the operator products appearing in the master equation equals the product of their time averages. Of course, this is not generally the case. For example, the time average of a product of functions has the following closed expression
\be\label{timeaverage2}
	\overline{F(t)G(t)} = \overline{F}(t) \overline{G}(t) +\sum_{n=1}^\infty\frac{\dt^{2n}}{2^nn!}\frac{d^n\overline{F}(t) }{dt^n} \frac{d^n\overline{G}(t) }{dt^n}\,,
\ee
 when a Gaussian windown function is considered.
If the time variation of $F,G$ is slow compared to $\delta t$, with characteristic frequency $\Omega$, then the average of the product is the product of the averages up to corrections of order ${\cal O}\left((\Omega \dt)^2\right)$. However, when the functions $F,G$ both have fast oscillatory behaviours characterized by a frequency $\w$ those corrections can add up to an exponentially large factor and then the leading term in the right hand side of (\ref{timeaverage2}) will not be a good approximation to its left hand side. Consider for example the case $F=e^{i\w t}F_0$ and $G= e^{-i\w t}G_0$ where $F_0$ and $G_0$ are constants.
} for $\rho$ takes the form:
\be
	i\partial_t\overline{\rho(t)} = [\overline{H}_{eff},\overline{\rho(t)}]+ i \left\{ \overline{A}, \overline{\rho(t)} \right\} + \overline{\gamma(t)}\ .\label{eq:timeaveqn}
\ee
The time-averaged operators are most easily written in the basis of eigenstates of $H_{IR}$, and are:
\begin{eqnarray}
%\overline{L(t)}& =& -\sum_{u\neq\bu}\sum_{ij}\frac{\langle u,i|V|\bu,j\rangle}{(\Delta E_u+\Delta E_{ij})}\langle \bu |V| u\rangle |i\rangle\langle j| \label{eq:Loperator}\\
\overline{{H}^{(2)}}&=&-\frac12\sum_{u\neq\bu}\sum_{ij}\left[\frac{\langle u,i|V|\bu,j\rangle}{(E_{u\bar{u}}+E_{ij})}\langle \bu |V| u\rangle |i\rangle\langle j|\right.\nonumber\\
& & \ \ \ \ \ \ \ \ \ \ \ \ \ \ \ \left.+\frac{\langle \bu,i|V|u,j\rangle}{(E_{u\bar{u}}-E_{ij})}|i\rangle\langle j| \langle u |V| \bu\rangle \right]\nonumber\\ \label{eq:avgham} \\
\overline{{A}^{(2)}}&=&-\frac12\sum_{u\neq\bu}\sum_{ij}\left[\frac{\langle u,i|V|\bu,j\rangle}{(E_{u\bar{u}}+ E_{ij})}\langle \bu |V| u\rangle |i\rangle\langle j|\right.\nonumber\\
& & \ \ \ \ \ \ \ \ \ \ \ \ \ \ \ \left.-\frac{\langle \bu,i|V|u,j\rangle}{(E_{u\bar{u}}- E_{ij})}|i\rangle\langle j| \langle u |V| \bu\rangle \right]\nonumber\\ \\
\overline{{\gamma}^{(2)}}&=&\sum_{u\neq\bu}\sum_{ij}\left[\frac{\langle u,i|V|\bu,j\rangle}{(E_{u\bar{u}}+ E_{ij})} |i\rangle\langle j|\overline{\rho^{(0)}} \langle \bu |V| u\rangle  
\right.\nonumber\\
& & \ \ \ \ \ \ \ \ \ \ \ \ \ \ \ \left.-\frac{\langle \bu,i|V|u,j\rangle}{(E_{u\bar{u}}- E_{ij})} \langle u |V| \bu\rangle \overline{\rho^{(0)}}|i\rangle\langle j|\right]\nonumber\\ \label{eq:gammaop}
\end{eqnarray}

\vskip .2cm
\noindent{\it Effective theories via Born-Oppenheimer expansion}.  These operators can be written in the form (\ref{eq:tlmaster},\ref{eq:tlham}).  Let us choose $k,l = um$ with $(u \neq {\bar u}$,  $m \in \{1,2\})$,  and $h_{um, u'm'} =  \delta(u,u')  |m-m'|$.  Then define
\begin{eqnarray}
	{\bar L}_{u,1} & = & \bra{u} V \ket{\bu} \nonumber\\
	{\bar L}_{u,2} & = & - i \sum_{ij} \frac{\ket{i}\bra{i}\bra{u} V \ket{\bu}\ket{j}\bra{j}}{E_{u,\bar{u}} + E_{ij}}\ , \label{eq:avglindops}
\end{eqnarray}
It is then easy to show that (\ref{eq:tlocevol},\ref{eq:tlmaster}) reduces to (\ref{eq:timeaveqn})\ with the operators defined as in (\ref{eq:avgham}-\ref{eq:gammaop}). We can write ${\bar L}_{u,2}$ in a more basis-independent form by expanding the denominator in a power series in $(E_{ij}/E_{u\bu})$ and noting that
$E_{ij} \ket{i} O_{ij} \bra{j} = \left[H_{IR}, \ket{i} O_{ij} \bra{j}\right]$:
\be
	{\bar L}_{u,2} = - i \frac{\bra{u} V \ket{\bu}}{E_{u\bu}} - i \sum_{k = 1}^{\infty} \frac{[\ldots[V_u,\overbrace{H_{IR}],\ldots, H_{IR}]}^{k\ times}}{E_{u{\bar u}}^{k+1}}
\ee
where $V_{u} = \bra{u} V \ket{\bu}$ is an operator acting on $\HIR$. The expansion in IR operators of increasingly high dimension weighted by inverse powers of $ E_{UV}$ is what we would expect from a good effective field theory, and is a central consequence of the hierarchical nature of the spectrum of the full (unreduced) theory.

\vskip .2cm
\noindent{\bf Leading order approximation}.  The master equation to the leading order in $\Delta E_{IR}/\Delta E_{UV}$  is:
\begin{eqnarray}
\overline{{H}^{(2)}}&=&-\sum_{u\neq\bu}\frac{V^{\dagger}_uV_u }{ E_{u \bu }}  - \half \sum_{u\neq\bu}\left[\frac{V^{\dagger}_u [V_u, H_{IR}]-[V^{\dagger}_u,H_{IR}]V_u}{ (E_{u\bu})^2}\right] + {\cal O}\left(\frac{1}{E_{u\bu}^3}\right)\nonumber \\
\overline{{A}^{(2)}}&=& -\frac12\sum_{u\neq\bu}\frac{[V^{\dagger}_u V_u, H_{IR}]}{ E_{u \bu }^2}  + {\cal O}\left(\frac{1}{E_{u\bu}^3}\right)\nonumber\\
\overline{{\gamma}^{(2)}}&=&\sum_{u\neq\bu}\left[\frac{[V_u, H_{IR}]\rho^{(0)}V_u^{\dagger}+V_u\rho^{(0)}[V^{\dagger}_u,H_{IR}]}{ E_{u \bu }^2}\right] 
+ {\cal O}\left(\frac{1}{E_{u\bu}^3}\right)
\end{eqnarray}
To leading order in $1/E_{u\bu}$, the evolution of ${\overline \rho}$ is completely Hamiltonian.  This is consistent with our discussion of the Born-Oppenheimer approximation in Appendix \S{A.3}, and with the results of \cite{2014AnPhy.345..141D,2014arXiv1405.2077D}. Our results display the kind of decoupling that occurs in Wilsonian renormalization: the effects of transitions to excited states of the UV degrees of freedom are suppressed by powers of $1/E_{UV}$.

One may ask whether the time-averaging we have implemented leads to a Markovian master equation. Once again, this will not happen beyond the leading order in $E_{IR}/E_{UV}$ for which the evolution is Hamiltonian.  If we consider the restricted case  $[H_{IR}, V_u] = a_u V_u$, $a_u \in \RR$, we find  ${\overline A}^{(2)} = {\overline \gamma}^{(2)} = 0$, so that the evolution is not only Markovian but Hamiltonian. Outside of this approximation, ${\bar L}_{u2}$ at order ${\cal O}(1/E_{u\bu}^2)$ is not proportional to $L_{u1}$, so that the negative eigenvalue of $h_{ui,uj}$ will contribute to (\ref{eq:tlmaster}).  To go further we must examine more specific cases.

\subsection{Examples}

We will work through two simple quantum-mechanical examples capturing our hierarchy of energy levels, in order to build up our intuition for the different possible dynamics of $\rho(t)$. In the first example of coupled spins, the non-Hamiltonian contributions will vanish upon time averaging. The second example is the well-studied case of coupled linear oscillators \cite{Caldeira:1982iu}; we will work with a different spectrum and quantum state for the ``bath", highlighting the differences between our results and those in \cite{Caldeira:1982iu}. Finally, we will give the time-averaged master equation for a scalar QFT with cubic self-coupling couplings, which to second order in perturbation theory can be computed with the formulae given.   This problem makes contact with the holographic setting of \cite{Balasubramanian:2012hb}. 

\subsubsection{Coupled spins}

First consider an IR spin coupled to $k=1\cdots M$ UV spins, all in the $2j+1$-dimensional irreducible representation of $SU(2)$ with spin $j$.  Thus the Hilbert space is  $\HIR = {\cal H}_{j_{IR}}$, $\HUV = \oplus_k {\cal H}_{j_{UV,k}}$. We take the Hamiltonian to be:
\begin{eqnarray}
	H_{IR} & = & -\mu_{IR} BS_{IR}^z\nonumber\\
	H_{UV} & = & -\sum_{k=1}^M\mu_{UV,k} B S_{UV,k}^z\nonumber\\
	\lambda V & = & \lambda\vec{S}_{IR}\cdot \sum_{k=1}^M\vec{S}_{UV,k}
\end{eqnarray}
where ${\vec S}$ are the usual spin operators, satisfying $[S_i, S_j] = i \hbar \epsilon_{ijk} S_k$ and $B$ is a fixed constant (a magnetic field).    We take $\mu_{IR} \ll \mu_{UV}$, so that this system has the hierarchical structure of energy levels we discussed above. 

We can rewrite the interaction term as
\be
	\lambda V = \lambda \sum_k S_{IR}^z S_{UV,k}^z +  \frac{\lambda}{2}\left(S_{IR}^-\sum_kS_{UV,k}^++S_{IR}^+\sum_kS_{UV,k}^-\right)
\ee
where $S_{\pm} = S_x \pm i S_y$ are the raising and lowering operators in the basis of $S_z$-eigenstates.  We write states in the basis
$\ket{j_{IR}, m_{IR}} \prod_k \ket{j_{UV,k} m_{UV,k}}$ where $j$ is the total angular momentum and $m$ the eigenvalue of $S_z$.
It is straightforward to see that the ground state of $H = H_{IR} + H_{UV} + \lambda V$ is independent of $\lambda$ to all orders in perturbation theory:
\be
	\ket{0} = \ket{j_{IR},m = j_{IR}} \prod_k \ket{j_{UV}, m = j_{UV}}
\ee
Thus, it is natural to consider an initial state of the form
\be
	\ket{\psi(0)} = C_m (S_{-,IR})^{j_{IR} - m} \ket{0} = \ket{j_{IR},m} \prod_k \ket{j_{UV}, m = j_{UV}}
\ee
which results from perturbing the ground state by an action of the operator $(S^-_{IR})^{j-m}$.

The terms in the  non-time-averaged equation of motion (\ref{eq:tlocevol}) to second order are:
\begin{eqnarray}
	H_{eff} & = & H_{IR} - \lambda\hbar  \left(\sum_k g_k j_k\right) S_{IR}^z \nonumber\\
	& & \ \ \ \ \ - \sum_k \frac{2 \lambda^2 \hbar^2 g_k^2 j_k}{(\mu_{UV,k} - \mu_{IR})B} \, \left(1 - \cos\left[(\mu_{UV,k} - \mu_{IR})B t\right]\right) \, S^-_{IR} \ , S^+_{IR} \nonumber\\
	A^{(2)} & = & - \sum_k \frac{\lambda^2 \hbar^2 g_k^2 j_k}{2 (\mu_{UV,k} - \mu_{IR})B} \, \sin \left[(\mu_{UV,k} - \mu_{IR})B t\right]	\, S^-_{IR} \,  S^+_{IR}\nonumber\\
	\gamma^{(2)} & = & \sum_k \frac{2 i \lambda^2 \hbar^2 g_k^2 j_k}{(\mu_{UV,k} - \mu_{IR})B} \, \sin \left[(\mu_{UV,k} - \mu_{IR})B t\right] \, S^+_{IR} \,  \rho^{(0)}(t) \, S^-_{IR}
\end{eqnarray}
where $\rho^{(0)}$ is the density matrix of the initial IR state evolved in time by the IR Hamiltonian $H_{IR}$.

$A,\gamma$ can be written in the form (\ref{eq:tlmaster})\ if the indices are expanded to $(k m)$ where $k$ labels the UV oscillators and $m\in \{1,2\}$.   With this notation,
\begin{eqnarray}
	h_{km,lm'} & = & \delta_{kl}  \, | m - m'|
	%\left(\begin{array}{ll} 0 & \frac{1}{\tau^*} \\ \frac{1}{\tau} & 0 \end{array}\right)_{ij}
	\nonumber\\
	L_{k1} & = & \hbar \sqrt{\frac{j_k}{2}} S^+_{IR} \nonumber\\
	L_{k2} & = & \hbar \sqrt{2j_k}\ \frac{\exp\left\{- \frac{i\left(\mu_{UV,k} - \mu_{IR}\right)B t}{2}\right\} \sin \left[\frac{\left(\mu_{UV,k} - \mu_{IR}\right)Bt}{2}\right]}{(\mu_{UV,k} - \mu_{IR})B} S^+_{IR}
\end{eqnarray}

In this example, the non-Hamiltonian terms $A^{(2)},\gamma^{(2)}$ are rapidly oscillating, and vanish after time averaging. 
%The discussion in the last paragraph of \S2.4 is precisely applicable in this case. We are starting in the ground state of the theory, so that $\ket{u}$ can only be reached by lowering operators, and $V_u \propto S^+_{IR}$.  Since $[S^z_{IR},S^+_{IR}] = S^+_{IR}$; the leading non-Hamiltonian term, and indeed the subleading ones, all vanish.   
For  completeness we compute the effective Hamiltonian for the time-averaged equation at ${\cal O}(\lambda^2)$:
\begin{eqnarray}
	{\overline H}_{eff} & = & H_{IR} - \lambda\hbar  \left(\sum_k g_k j_k\right) S_{IR}^z - \frac{2 \lambda^2 \hbar j}{B} \left[\sum_k 
		\frac{g_k^2}{\mu_k - \mu_{IR}} \right] S^-_{IR} S^+_{IR} + {\cal O}(\lambda^3)\nonumber\\
		& = & - {\tilde \mu} B S^z_{IR} + \beta (S^z_{IR})^2 - E_g
\end{eqnarray}
where
\begin{eqnarray}
	\tilde \mu & = & \mu_{IR} - \frac{\hbar \lambda}{B} \sum_k g_k j_k - \frac{\hbar^2 \lambda^2}{B}\sum_k \frac{g_k^2 j_k}{\mu_k - \mu_{IR}} \nonumber\\
	\beta & = & \frac{\hbar \lambda^2}{B}\sum_k \frac{g_k^2 j_k}{\mu_k - \mu_{IR}}\nonumber\\
	E_g & = & - \frac{\hbar^3 j(j+1)}{B} \sum_k \frac{g_k^2 j_k}{\mu_k - \mu_{IR}}
\end{eqnarray}
$H_{eff}$ is related to $H_{IR}$ by renormalization of the magnetic moment, coupling $\beta$, and vacuum energy.	

%For non-factorized initial states of the form
%%
%$$\ket{\Psi} = \frac{1}{\sqrt{2}} \left( \ket{\psi_{1,IR}}\ket{j_{UV},m_1} + \ket{\psi_{2,IR}}\ket{j_{UV},m_2}\right)$$
%%
%the first-order corrections to the time-averaged evolution equations vanish.  We will leave this example for future work.

\subsubsection{Linear oscillators}

Following  \cite{Caldeira:1982iu,Leggett:1987zz}, we consider $\HIR$ the Hilbert space of a simple harmonic oscillator, and $\HUV$ a bath of harmonic oscillators, with the Hamiltonian comprising a linear coupling between them:
\begin{eqnarray}
	H_{IR} & = & \frac{P^2}{2M} + \half M \Omega^2 X^2 \nonumber\\
	H_{UV} & = & \sum_k \left[ \frac{p_k^2}{2 m_k} + \half m_k \omega_k^2 x_k^2\right]\nonumber\\
	\lambda V & = & \sum_k C_k x_k X\label{eq:coupledoscmodel}
\end{eqnarray}
where we take $C_l \sim {\cal O}(\lambda)$.   Of course, this can be solved exactly by a change of variables.  However, in the spirit of this paper we are interested in the dynamics of the ``bare" variable $X$, to which we imagine our measuring devices couple.

In order to match what we expect from a quantum field theory calculation such as that outlined in the introduction, we will take the frequencies $\omega_k \gg \Omega$. Furthermore, we will assume that the UV oscillators are in an eigenstate of $H_{UV}$ (such as the ground state) at leading order in perturbation theory.  The resulting system then differs from those studied in \cite{Caldeira:1982iu,Leggett:1987zz}. Those works consider the oscillators $x_k$ to be some ``environment", with a spectrum designed phenomenologically to model quantum Brownian motion or dissipation. The environment contains oscillators with arbitrary low frequency, to model dissipation of energy and phase coherence into an environment, over time scales long compared to a given experiment.  Furthermore, we are most interested in the UV oscillators initially in their ground state -- thus, our treatment is closest to the zero-temperature limit of  \cite{Caldeira:1982iu,Leggett:1987zz}.  In this case, some approximations made in those works fail.  Finally, we implement time averaging differently, by directly averaging the density matrix over a coarse-graining kernel.  The net result is a qualitatively different master equation for the density matrix.

As stated, we assume that at $t = 0$, the UV oscillators are in an energy eigenstate $\ket{{\bar u}} = \prod_{\ell} \ket{n_{\ell}}$.  In this case, the first-order shift of the Hamiltonian vanishes, because the expectation value of $x_{\ell}$ vanishes in energy eigenstates of the harmonic oscillator. 
% At order $\lambda^2$, we could do the computation starting with (\ref{eq:avglindops}). A more interesting expression arises from computing $H, A,\gamma$ and extracting a new representation of the form (\ref{eq:tlmaster}).
Using Eqs. (\ref{eq:avgham}) -- (\ref{eq:gammaop}) we find that
the operators in the second-order time-averaged equation (\ref{eq:timeaveqn}) are:
\begin{eqnarray}
	{\overline H}^{(2)} & = & - \frac{1}{2}\sum_{\ell} \frac{(2 n_{\ell} + 1) C_{\ell}^2}{ m_{\ell} (\omega_{\ell}^2 - \Omega^2)} \left[  X^2 -\frac{\hbar}{2M \omega_{\ell}}\right]\nonumber\\
	{\overline A}^{(2)} & = & -\frac{1}{2} \sum_{\ell} \frac{(2 n_{\ell} + 1) C_{\ell}^2}{2 M m_{\ell} \omega_{\ell} (\omega_{\ell}^2 - \Omega^2)} \left\{X,P\right\}\nonumber\\
	{\overline \gamma}^{(2)} & = & i \sum_{\ell} \frac{(2 n_{\ell} + 1) C_{\ell}^2}{2 M m_{\ell} \omega_{\ell} (\omega_{\ell}^2 - \Omega^2)} \left(P {\overline \rho}^{(0)}(t) X + X {\overline \rho}^{(0)}(t) P \right)
\end{eqnarray}
At this order, the Hamiltonian is changed by a shift in the oscillator frequency and the ground state energy.  We can rewrite $A,\gamma$ in the form (\ref{eq:tlmaster}) if we let $m$ run from 1 to 2 and define:
\begin{eqnarray}
	h_{12} & = & h_{21} = \sum_{\ell} \frac{(2n_{\ell} + 1) C_{\ell}^2}{2 m_{\ell} (\omega_{\ell}^2 - \Omega^2)} \nonumber\\
	h_{11} &=& h_{22} = 0 \\
	L_1 & = & X\nonumber\\
	L_2 & = & -i\left(X+i\frac{P}{M \omega_{\ell}}\right)
\end{eqnarray}
Thus the eigenvalues of the $h_{ij}$ matrix are $\pm h_{12}$.  As we explained before, the lack of positive  definiteness implies that the evolution of $\rho$ is not Markovian beyond the leading order in $\Omega/\omega_{\ell}$.

Before continuing, it is worth comparing the form of our master equation to that of Caldeira and Leggett \cite{Caldeira:1982iu}.  This arises when 
\begin{itemize}
\item $x_{\ell}$ describes a continuous spectrum of oscillators, for which 
\be \sum_{l} C_{\ell}^2 f(\omega_{\ell}) \to \int d\omega \rho(\omega) C(\omega)^2 f(\omega)\ ,
\ee
and $m_{\ell} = m$, with 
\be
	\rho C^2(\omega) = \frac{2 m \eta \omega^2}{\pi} \theta(\Lambda - \omega)
\ee
where $\Lambda$ is some UV cutoff, and $\eta$ a phenomenologically determined coefficient.
\item Furthermore, the oscillators $x_{\ell}$ are placed at finite temperature $T \gg \Lambda$.
\end{itemize}
In this case they derive a master equation (Equation (5.12) in \cite{Caldeira:1982iu}) which can be rewritten in the form (\ref{eq:tlocevol},\ref{eq:tlmaster})\ with the indices $i,j \in \{1,2\}$ and\footnote{In fact the master equation in \cite{Caldeira:1982iu}, which yields the operators $L_i$ we report, is missing a term of order $kT/\Lambda$; this term is argued to be small even in the $kT \gg \Lambda$ limit. We discuss this further in \cite{Agon:2017oia}.}
\begin{eqnarray}
	h_{12} & = & h_{21} = \frac{\eta \Lambda }{2 \hbar} \nonumber\\
	L_1 & = & X \nonumber\\
	L_2 & = & -i\left(X -\frac{\hbar }{2 M \Lambda } P \right) + \frac{2kT}{\Lambda}X
	\label{eq:clmaster}\\
\end{eqnarray}
where $\eta$ is a function of $C(\omega)$ and the UV cutoff, and $\Lambda$ is a UV energy scale that accounts for the frequency renormalization. Note the relative factor of $-i$ in the coefficient of $P$, as well as the additional temperature dependent term proportional to $X$ in $L_2$.  In general, their master equation is also not Markovian, unless we were to take the limit $kT \to \infty$, $\eta k T$ finite. ({\it cf.}\ \cite{breuer2007theory}). 

As discussed in \cite{Agon:2017oia}, this model captures some essential features of local quantum field theories, if we choose $C(\omega)$ appropriately.  In particular, there can be divergences when the number of states grows sufficiently rapidly with energy.  

\subsubsection{Scalar QFT with cubic self-coupling}

At second order in perturbation theory, it is straightforward to apply our formulae to scalar quantum field theories.    We give a brief description here of the the cubic theory in $d$ spatial dimensions. A fuller account of our computation, and interpretation of the resulting divergences, can be found in \cite{Agon:2017oia}. Here our goal is to demonstrate features of the master equation also found in a holographic context in \cite{Balasubramanian:2012hb}. 

Consider the Lagrangian
\be\label{eq:scalarlag}
	{\cal L} = \half (\p\phi)^2 - \half m^2 \phi^2 - \frac{g}{3!} \phi^3
\ee
The master equation for the quartic theory in four dimensions was computed in \cite{Lombardo:1995fg}, via the Feynman-Vernon influence functional.  We wish to consider (\ref{eq:scalarlag}) in light of our more abstract conceptualization; in addition, with our Hamiltonian regulator, we will find some dimension-dependent issues that would be absent in four dimensions.

The scalar field in the interaction picture can be decomposed as follows:
\begin{eqnarray}
	\phi (x,t) & = & \int_{|{\vec k}| < \Lambda} \frac{d^d k}{\sqrt{(2\pi)^d 2 \omega(k)}} \left\{a_{IR}(k) e^{i \vk \cdot \vx}  + a^{\dagger}_{IR}(k) e^{-i \vk\cdot\vx} \right\}\nonumber\\
	& & + \qquad\qquad \int_{M > |{\vec k}| > \Lambda} \frac{d^d k}{\sqrt{(2\pi)^d 2 \omega(k)}} \left\{a_{UV}(k) e^{i \vk \cdot \vx}  + a^{\dagger}_{UV}(k) e^{-i \vk\cdot\vx} \right\}\nonumber\\
	& = & \phi_{IR}(x,t) + \phi_{UV}(x,t) \label{eq:scalarcoordft}
\end{eqnarray}
where $\omega(k) = \sqrt{\vk^2 + m^2}$, $\Lambda$ is the spatial coarse-graining scale, and $M$ is the cutoff. 
The dual conjugate momentum is
\begin{eqnarray}
	\pi(x,t) & = &  \int_{|{\vec k}| < \Lambda} \frac{d^d k (-i \omega(k))}{\sqrt{(2\pi)^d 2 \omega(k)}} \left\{a_{IR}(k) e^{i \vk \cdot \vx}  - a^{\dagger}_{IR}(k) e^{-i \vk\cdot\vx} \right\}\nonumber\\
	& & +  \qquad\qquad \int_{M > |{\vec k}| > \Lambda} \frac{d^d k (-i \omega(k))}{\sqrt{(2\pi)^d 2 \omega(k)}} \left\{a_{UV}(k) e^{i \vk \cdot \vx}  - a^{\dagger}_{UV}(k) e^{-i \vk\cdot\vx} \right\}\nonumber\\
	& = & \pi_{IR}(x,t) + \pi_{UV}(x,t) \label{eq:scalarmomft}
\end{eqnarray}
The normalized single-particle momentum eigenstates are $|k\rangle=\sqrt{2 \w_k}a^{\dagger}_k|0\rangle$, where 
$\left[a_k,a^\dagger_{k'}\right]=\delta^{d}(k-k')$.

There is thus a decomposition of the Hilbert space 
\be
	\CH = \CH_{IR} \otimes \CH_{UV}
\ee
We can split the Hamiltonian accordingly into $H = H_{IR} + H_{UV} + g V$ where
\begin{eqnarray}
	H_{IR,UV} & = &  \half (\p\phi_{IR,UV})^2 - \half m^2 \phi_{IR,UV}^2 - \frac{g}{3!} \phi_{IR,UV}^3\nonumber\\
	g V & = & \frac{g}{2} \phi_{IR} \phi_{UV}^2 + \frac{g}{2} \phi_{IR}^2 \phi_{UV}
\end{eqnarray}
In essence, each oscillator with UV momentum acts as a separate harmonic oscillator.

We will take the initial state to be of the form $\ket{\Psi} = \ket{\psi}_{IR} \ket{0}_{UV}$, where $\ket{0}_{UV}$ is the vacuum with respect to $H_{IR}$. With $g V$ defined as above there are two classes of matrix elements that contribute to (\ref{eq:notimeav},\ref{eq:avgham}-\ref{eq:gammaop}), i.e. to $H_{eff}$, $A$ and $\gamma$ that control the time evolution of the IR density matrix:
\begin{enumerate}
\item Creation of a single particle in $\HUV$. This means that the relevant components of $\phi_{IR}^2$ will be two nearly collinear particles in $\HIR$ with total momentum $\vk_{1,IR} + \vk_{2,IR} = \vk_{UV}$,. Both IR and UV momenta must have magnitudes close to the scale $\Lambda$ that splits IR from UV.
\item Creation of two particles in $\HUV$.  The matrix elements that contribute will have two excitations in $\HUV$ with nearly back-to-back momenta that sum to momentum with magnitude below $\Lambda$.  There is a much larger set of possibilities: almost any magnitude of UV momentum will be allowed, and for each value $k$ there will be a sphere in phase space of volume $\sim k^{d-1}$ of possible UV momenta.
\end{enumerate}
The importance of each type of term depends on the IR momenta, and on the number of dimensions.  For low enough IR momenta, only the second type of term can contribute. For simplicity, we will focus on this possibility.

A straightforward application of our formalism yields
\begin{eqnarray}
	H^{(2)}(t)&=&-\frac i4 \int_0^t d\tau \int_{uv} \, \frac{d^dk}{2\w_k}\frac{d^dk'}{2\w_{k'}}\Big[\langle 0|V|k k'\rangle \, \langle k k'|V_I(-\tau)|0\rangle - h.c. \Big] \nonumber \\
	A^{(2)} & = & - \frac{1}{4} \int_0^t d\tau \int_{uv} \, \frac{d^dk}{2\w_k}\frac{d^dk'}{2\w_{k'}}\Big[\langle 0|V|k k'\rangle \, \langle k k'|V_I(-\tau)|0\rangle+ h.c. \Big] \nonumber \\
	\gamma^{(2)} & = & \frac{i}{2}\int_0^t d\tau \int_{uv} \,  \frac{d^dk}{2\w_k}\frac{d^dk'}{2\w_{k'}}\Big[\langle 0|V|k k'\rangle \, \rho^{(0)} \, \langle k k'|V_I(-\tau)|0\rangle + h.c. \Big]\label{eq:qftmast}
\end{eqnarray}

The integral of the time-dependent matrix element is:
\begin{eqnarray}
	& & \int_0^t d\tau \,_{uv}\langle k k'|V(-\tau)|0\rangle_{uv} =  \lambda\int \frac{d^dx}{(2\pi)^d}e^{-i(k+k')\cdot x}\int_0^t d\tau\phi_{ir}(-\tau)e^{-i(\w_k+\w_{k'})\tau}\nonumber\\
	& & \qquad =-i\lambda\int \frac{d^dx}{(2\pi)^d}e^{-i(k+k')\cdot x}\int_{ir} \frac{d^dp}{\sqrt{(2\pi)^d 2\w_p}}\left[\frac{a_p \left(1 - e^{i (\omega_p - \omega_k - \omega_{k'})t }\right)}{\w_k+\w_{k'} -\w_p}\right.\nonumber\\
	& & \qquad \qquad \qquad \qquad \qquad \qquad \qquad \qquad \left. +\frac{a^{\dagger}_{-p}\left(1 - e^{- i (\omega_p + \omega_k + \omega_{k'})t }\right)}{\w_k+\w_{k'}+\w_p}\right] e^{ipx}\nonumber\\ \label{eq:matrixel}
\end{eqnarray}
It is clear that $H,A,\gamma$ will have time dependence on scales of order the UV momenta, which range from $\Lambda$ to $M$.  

Next, let us consider the time averaged quantities ${\bar H}^{(2)}$ and ${\bar A}^{(2)}$. This amounts to dropping the rapidly oscillating exponential terms in (\ref{eq:matrixel})\ to find:
\begin{eqnarray}
	&&\int_0^t d\tau \int_{uv} \frac{d^dk}{2\w_k}\frac{d^dk'}{2\w_{k'}}\langle 0|V|k k'\rangle\langle k k'|V(-\tau)|0\rangle\longrightarrow \nonumber\\
&&-i\lambda^2\int \frac{d^dx}{(2\pi)^d}\phi_{ir}(x)\int_{ir} \frac{d^dp}{\sqrt{(2\pi)^d 2\w_p}}(a_p+a^{\dagger}_{-p})e^{ipx}\nonumber\\
&& \qquad \qquad \qquad \qquad \times \int_{uv}\frac{d^dk}{2\w_k}\frac{\w_k+\w_{{p - k}}}{2\w_{p-k}[(\w_k+\w_{p-k})^2-\w_p^2]} \nonumber \\
&&-\lambda^2 \int \frac{d^dx}{(2\pi)^d}\phi_{ir}(x)\int_{ir}\frac{d^dp\, i\w_p}{\sqrt{(2\pi)^d 2\w_p}}(a^{\dagger}_{-p}-a_p)e^{ipx}\nonumber\\
&&\qquad \qquad \qquad \qquad \times\int_{uv}\frac{d^dk}{2\w_k}\frac{1}{2\w_{p-k}[(\w_k+\w_{p-k})^2-\w_p^2]}%\nonumber\\
%& & = -i\lambda^2\int \frac{d^dx}{(2\pi)^d}\phi_{ir}(x)\int_{ir} \frac{d^dp}{\sqrt{(2\pi)^d 2\w_p}}(a_p+a^{\dagger}_{-p})e^{ipx} F\left({\vec p},\omega_p, \Lambda, M\right)\nonumber\\
%& & \ \ \ - \lambda^2 \int \frac{d^dx}{(2\pi)^d}\phi_{ir}(x)\int_{ir}\frac{d^dp\, i\w_p}{\sqrt{(2\pi)^d 2\w_p}}(a^{\dagger}_{-p}-a_p)e^{ipx} G\left({\vec p},\omega_p,\Lambda,M\right)\nonumber\\
\end{eqnarray}
Note that the second and third lines, proportional to $-i \lambda^2$, will contribute to $H$, while the fourth and fifth lines, proportional to $-\lambda^2$, will contribute to $A,\gamma$.

Let us first examine the correction to the Hamiltonian. If $|{\vec p}|, \omega_p \ll \Lambda \ll M$, we can expand the integral
\be
	\int_{uv}\frac{d^dk}{2\w_k}\frac{\w_k+\w_{{p - k}}}{2\w_{p-k}[(\w_k+\w_{p-k})^2-\w_p^2]}
	= F(M,\Lambda) \left(1  + h^d_1 \frac{{\vec p}^2}{\Lambda^2} + \ldots 
	\right)
\ee
where
\be
 F = F_0 \left\{ \begin{array}{l} \Lambda^{d-3} \qquad d < 3 \\ 
 \ln\left(\frac{M}{\Lambda}\right) \qquad d = 3 \\
 M^{d-3} \qquad d > 3 \end{array}\right.\label{eq:divergone}
\ee
and $F_0, h^d_1$ are dimensonless constants.  The neglected terms include both higher orders in ${\vec p}^2$ as well as terms suppressed by powers of $(\Lambda/M)^2$.  These terms give corrections to the Hamiltonian of the form
\be
	{\bar H}^{(2)} \propto \lambda^2 \int d^d x F(M,\Lambda) \left( \phi_0^2 + \frac{h^d_1}{\Lambda^2}\ \pi \vec{\nabla}^2\phi + \ldots\right)
\ee
where we the dots indicate terms of higher order in $\vec{\nabla}^2/M^2$, $\vec{\nabla}^2/\Lambda^2$.  It is thus clear that we will get terms with spatial nonlocalities on scales $\Lambda, M$. Note that for $d \geq 3$, we find divergent contributions to the mass, consistent with standard treatments of scalar QFT.

Next consider corrections to ${\bar A}^{(2)}$ In this case, 
\be
	\int_{uv}\frac{d^dk}{2\w_k}\frac{1}{2\w_{p-k}[(\w_k+\w_{p-k})^2-\w_p^2]} = G(M,\Lambda) \left(1  + {\tilde h}^d_1 \frac{{\vec p}^2}{\Lambda^2} + \ldots 
	\right)
\ee
where
\be
 G = G_0 \left\{ \begin{array}{l} \Lambda^{d-4} \qquad d < 4 \\ 
 \ln\left(\frac{M}{\Lambda}\right) \qquad d = 4 \\
 M^{d-3} \qquad d > 4 \end{array}\right.\label{eq:divergtwo}
\ee
and $G_0$, ${\tilde h}^d_1$ are dimensionless coefficients.  These will lead to corrections of the form
%'
\be
	{\bar A}^{(2)} \propto \lambda^2 \int d^d x G(M,\Lambda)\left(\phi \pi + \frac{{\tilde h}^d_1}{\Lambda^2}\ 
		\phi {\vec \nabla}^2 \pi + h.c. + \ldots\right)
\ee
Again, it is clear that the additional terms, higher order $\vec{\nabla}^2/M^2$, $\vec{\nabla}^2/\Lambda^2$, lead to spatial nonlocality on scales of order $\Lambda,M$.  The terms of the form $\{\pi,\phi\}$ are clearly analogous to the $\{X,P\}$ terms which appear in ${\bar A}^{(2)}$ for the linearly coupled oscillator.

We can see from Eqs. (\ref{eq:divergone},\ref{eq:divergtwo}) that we also have new divergences in high enough dimension.  In prior studies of quartic scalar field theory \cite{Lombardo:1995fg}, the non-Hamiltonian terms in the master equation (derived from the influence functional) had no divergences.  There are two differences here.  We look at more general spacetime dimensions; and we adopt a spatial regulator appropriate to our Hamiltonian treatment, after the fashion of \cite{Kogut:1974ag}. This is discussed in more depth in  \cite{Agon:2017oia}.

The upshot is that we have non-Markovian behavior, indicating the development of entanglement between the UV and IR, and nonlocality on the order of the cutoff $\Lambda$. This is precisely the structure hinted at in \cite{Balasubramanian:2012hb}. 

\section{Discussion and conclusions}
%\subsection{Some important next steps}

\subsection{Relation to holographic RG}

The results in \cite{Balasubramanian:2012hb}\ indicate that the ``IR region" of an AdS geometry, i.e. the interior region far from the spacetime boundary, functions as an open quantum system. However, following \cite{Heemskerk:2010hk,Faulkner:2010jy}, this work computed the Feynman path integral assuming vacuum boundary conditions in the far past and far future, integrating out degrees of freedom beyond some radial position.  As we note in Appendix A, this only makes sense if we have foreknowledge of the degrees of freedom we are integrating out. Such foreknowledge for interacting systems makes sense if either 
\begin{enumerate}
\item We are working with renormalized variables in which the Hamiltonian is block diagonal between low and high energies, and wish to only measure these redefined variables in the low-energy Hilbert space.
\item We are computing scattering amplitudes for asymptotic states of well-separated particles, 
\end{enumerate}
The supergravity modes inside a radial cutoff in AdS/CFT clearly do not correspond to the renormalized variables in point (1).  As for (2), in global AdS coordinates there are no good well-separated asymptotic states in AdS, as all excitations oscillate on times scales of order $R_{AdS}$.  More generally, the analog of S-matrix elements, for which an LSZ-type reduction applies, are correlators of local CFT operators, dual to non-normalizable modes supported near the AdS boundary \cite{Balasubramanian:1998sn,Balasubramanian:1999re,Balasubramanian:1999ri}.  

Nonetheless, we can already learn something from \cite{Balasubramanian:2012hb}. The first is that the holographic Wilsonian action is nonlocal in time.  Thus, this action cannot describe purely Hamiltonian dynamics, exactly as expected for an open quantum system.  The time scale which describes mixing between the IR and the UV is of order the cutoff.  This is exactly what we find for the density matrix dynamics in \S2.  The non-time averaged master equation has oscillations at time scales of order $\Delta E_{UV}$, which become time-independent upon time averaging. Similarly, UV-IR entanglement evolves on time scales of order $\Delta E_{UV}$.

As we will discuss below, for open quantum systems the correct analog of the Wilsonian action is the logarithm of the Feynman-Vernon influence functional, computed via path integral techniques as we describe in Appendix B.   Applying this approach to the AdS/CFT correspondence with a bulk radial cutoff faces the  challenge because the gauge theory interpretation of such a cutoff remains unclear \cite{Balasubramanian:2012hb,Balasubramanian:2013lsa,Mintun:2014gua}. However, in the limit that we can study small quantum fluctuations in anti-de Sitter space, it would be of great interest to compute the dynamics of a density matrix for scalar fields supported in the region $r < r_{\Lambda}$ in AdS spacetime, dual to scalar operators on the boundary.

%\cite{Balasubramanian:2013lsa,Mintun:2014gua}
%
%We have given a very preliminary description of coarse graining in a set of model systems with a particularly simple structure of energy levels, a simple band structure determined by two scales $\Delta E_{UV}, \Delta E_{IR}$.  A local quantum field theory, however, will have a nested hierarchy of energy levels, determined by a hierarchy of scales $\Delta E_1 \ll \Delta E_2 \ll \Delta E_3 \cdots \ll \Delta E_N$.  In perturbation theory, one would first trace out the bare degrees of freedom characterized by energy scale $\Delta E_N$, and so on down. We expect that the terms $H, A, \gamma$ in the master equation for $\rho$ will run smoothly under this successive integrating out, just as they do in prior treatments of Hamiltonian renormalization. More generally, it will be important to explicitly compute the master equation for various quantum field theories.

\subsection{Coarse graining in the path integral}

%For open quantum systems, the correct object to coarse grain is the density matrix, with a master equation 

Because we are computing time-dependent inclusive probabilities for measurements of operators supported in the IR, the correct object to coarse-grain is not a transition amplitude but the density matrix for the system.  The computation of the density matrix in path integral language goes back to Feynman and Vernon \cite{Feynman:1963fq}.  Let us state the essential formulae here; the derivation, and a perturbation theory calculation relevant for \S2.3, can be found in Apppendix D.  

We assume following \cite{Feynman:1963fq}\ that the density matrix at time $t=0$ is factorized between the IR and the UV: that is, there is no initial entanglement. Let capital letters $X,Y$ denote quantum-mechanical variables describing the IR, and lower case letters $x,y$ denote quantum variables describing the UV. We assume the action can be written as 
\be
	S[X,x] = S_{IR}[{\dot X},X] + S_{UV}[{\dot x}, x] + \delta S[x, X]
\ee
The corresponding Hamiltonian will take the form (\ref{eq:splithamiltonian}). Note that this is a somewhat restrictive action; in particular, the interactions involve coordinates and not velocities.

The density matrix for the IR degrees of freedom $X$, as a function of time, can be written as:
\begin{eqnarray}
\rho(X,Y,t)&=&\int dX' dY' J(X,Y,t;X'Y',0)\rho_{init}(X',Y',0)\,, \nonumber\\
J(X,Y,t;X'Y',0)&=&\int_{Y(0)=Y';\,X(0)=X'}^{Y(t)=Y;\,X(t)=X}{\cal D} X {\cal D} Y e^{\frac i{\hbar}S_{IR}[X]-\frac i{\hbar}S_{IR}[Y]}{\cal F}[X(t),Y(t)]\,,\nonumber\\ \label{eq:fpidensity}
\end{eqnarray}
where $\rho_{init}$ is the initial density matrix for the IR degrees of freedom, and the {\it influence functional} 
\begin{eqnarray}
	\CF[X(t),Y(t)]&=&\int dx' dy' dx \rho_{_U}(x',y',0) \nonumber \\ 
	&\times &\int_{y(0)=y';\,x(0)=x'}^{y(t)=x;\,x(t)=x} {\cal D}x{\cal D}y e^{\frac i{\hbar}S_U[x]-\frac i{\hbar}S_U[y]+\frac i{\hbar}\delta S[x,X]-\frac i{\hbar}\delta S[y,Y]}\nonumber\\ \label{eq:influencedef}
\end{eqnarray}
contains the dependence on the initial state of the UV degrees of freedom, as well as the interactions between the UV and IR degrees of freedom. In general this cannot be written in the form $F[X]G[Y]$. The influence functional encodes the same data as the terms $H_{eff},A,\gamma$ in the master equation (see Appendix D)

The point of stating these well-known results is to emphasize that the correct analog of the Wilsonian effective action, in the case that the IR and UV are entangled and one is asking questions about finite-time processes, is the influence functional. This point has been made eloquently in a number of papers, including \cite{Lombardo:1995fg,Calzetta:1996sy,Calzetta:1999zr}.   However, this approach has not been applied to holographic Wilsonian renormalization.  It should be.

The Wilsonian approach to renormalizing the path integral is based on the Euclidean path integral, and the coarse-graining is over Euclidean space (that is, the cutoff is placed on Euclidean momenta).  In a real-time context this cutoff makes less sense, and indeed \cite{Lombardo:1995fg,Calzetta:1996sy,Calzetta:1998ng,Calzetta:1999zr} coarse-grain with respect to spatial momenta.  However, in most physical processes we will also have finite accuracy in determining the times at which we prepare and measure the system, and temporal coarse-graining is required.  It would be of great interest to find a simple path integral implementation of the time averaging that we discussed in this paper.

%isolate the contributions of $H_{eff}^{(2)}$, $A_{eff}^{(2)}$, and $\gamma^{(2)}$.  It is clear that
%%
%\be
%	L^{(2)}(t) = H^{(2)}_{eff} + i A^{(2)}_{eff} = -\frac{ i \lambda ^2}{\hbar} \sum_{a,b} \int_0^t  dt' G^F_{c; a,b}(t,t') \Phi_{a}[t]\Phi_{b}[t] 
%\ee
%%
%so that
%%
%\begin{eqnarray}
%	H_{eff}^{(2)} & = & \frac{2 \lambda^2}{\hbar}\sum_{a,b} \int^t_0 dt' \Ima G^F_{c;a,b}(t,t') \Phi_a(t) \Phi_b(t)\nonumber\\
%	A_{eff}^{(2)} & = & - \frac{2 \lambda^2}{\hbar}\sum_{a,b} \int^t_0 dt' \Rea G^F_{c;a,b}(t,t') \Phi_a(t) \Phi_b(t)\nonumber\\
%	\gamma^{(2)} & = & \frac{i \lambda^2}{\hbar} \sum_{a,b} \int_0^t dt' \left(G^W_{c;ab}(t,t') + G^W_{c;ab}(t',t)\right)
%		\Phi_a(t) \rho_{_I}^{(0)}(t) \Phi_b(t)
%\end{eqnarray}
%%
%A bit of algebra\footnote{Check again!} shows that these expressions are identical to those found in \S3.

%One technical step we have not achieved is a nice implementation of finite temporal resolution directly in the path integral.  
%We reiterate that the non-Ha
%
%\begin{itemize}
%\item Return to the point that the standard questions people ask involve low-energy states that are dressed and factorized from the UV in that sense.  So they look  Hamiltonian (and thus Markovian) in the effective  dynamics
%\item We did not get Markovian dynamics amongst other things because we don't necessarily satisfy the above conditions.
%\end{itemize}
%

\subsection{Additional questions}

\noindent{\it Strongly coupled systems}. In using the phrase ``coarse-graining" in our perturbative treatment, we implied that we could assign an energy scale to the degrees of freedom we were tracing out, corresponding to a short distance scale.  This makes sense in weakly-coupled quantum field theories, in which the energy and momenta of single quanta are tied together and there is some meaning to these single quanta.   In strongly-coupled systems, especially those without quasiparticle excitations, this relation breaks down.  It would be interesting to study our coarse-graining in such examples, either analytically or numerically.

\vskip .2cm

\noindent{\it Cosmological perturbations}.  Primordial non-gaussianities in CMB fluctuations and large-scale structure measure correlations between quantum fluctuations at different scales, induced by interactions in the inflaton sector.  We expect the initial state to have a degree of quantum entanglement between scales, following the discussions in this paper, which should help seed the classical correlations one actually observes.

Some discussion of entanglement between scales during inflation appears in \cite{Burgess:2014eoa}.  In this work, the entanglement between short- and long-wavelength modes is used to justify a Lindblad equation describing Markovian evolution for the long-wavelength modes, based on an argument that the Hubble scale sets a natural time scale for the decay of correlations of short-wavelength modes.  It would be interesting to perform a more quantitative, first-principles analysis of entanglement between scales in some specific model, following the discussion here.  For example, as we have noted, even when correlation functions are essentially local in time, the dynamics of long-wavelength modes can still fail to be Markovian.

\vskip .2cm

\noindent{\it Holographic renormalization}.  In our setup, the evolution equation for $\rho$ is local on scales larger than $\Delta E_{UV}$.  This is in accord with the discussion of holographic gauge theories in \cite{Heemskerk:2010hk,Faulkner:2010jy,Balasubramanian:2012hb,Mintun:2014gua}, in which the Wilsonian effective action of a strongly-coupled field theory was nonlocal on the scale of the cutoff, reflecting the propagation of excitations into and back out of the UV region.  However, string theory suggests that there are other nongravitational theories in which the time scale over which excitations are supported in the UV becomes arbitrarily large.  One example is little string theory -- in the holographic dual, massless excitations propagating into the UV region take an infinite time to reach the ``boundary".  This is tied to the exponential (Hagedorn) growth of states at high energies in this theory.  It would be interesting to explore the dynamics of $\rho_{IR}$ in this setting.

More generally we would like a more precise understanding of the relationship between the framework in this paper and that of Wilsonian renormalization in holographic gauge theories, in which one ``integrates out" a section of the geometry \cite{Balasubramanian:1999jd,Heemskerk:2010hk,Faulkner:2010jy,Balasubramanian:2012hb}.  For example, this could cast an interesting light on black hole entropy.  There is evidence that bulk quantum corrections to the entanglement entropy of quantum fields between the interior and exterior of a ``stretched horizon" outside the black hole are mapped to the Wald entropy of the black hole, using the renormalized gravitational action (see \cite{Jacobson:2012ek,Cooperman:2013iqr}\ and references therein); and there are conjectures that the full Bekenstein-Hawking/Wald entropy of the black hole can be considered as an entanglement entropy (see for example \cite{Bianchi:2012ev}). 

In holographic theories, black holes are dual to high-energy states with thermal behavior.  In closed quantum systems, the Eigenstate Thermalization Hypothesis \cite{Deutsch:1991,Srednicki:1994}\ states that a class of quantum operators will have expectation values and correlation functions which appear to be thermal.  In many examples, these are local operators supported in a spatial subregion of the system, and the excited quantum state is strongly entangled between the subregion and its complement so that the reduced density matrix looks approximately thermal (see \cite{Nandkishore:2014kca}\ for a recent discussion, and further references.) Of course, this is not the only way to decompose the Hilbert space such that the state is entangled between the components.  The stretched horizon appears at some radius in the AdS-black hole geometry, whose value should be dual to some scale in the field theory dynamics.  Studying entanglement between degrees of freedom at different scales could shed light on this system.\footnote{On a related note, Ref. \cite{Jacobson:2012ek}\ studies the progressive contribution of longer and longer wavelengths of {\it bulk}\ fields to the black hole entanglement entropy.}

\vskip .3cm

\noindent{\bf Acknowledgments:} We thank Alberto Guijosa, Matthew Headrick, Robert Konick, Sergei Khlebnikov, Sung-Sik Lee, Hong Liu, Emil Martinec, Greg Moore, Vadim Oganesyan, Anatoli Polkovnikov, and Stephen Shenker for useful questions, conversations, and (in some cases) encouragement.   We thank Bei-Lok Hu for pointing us to \cite{fleming2011accuracy}, and Vladimir Rosenhaus for pointing out some mistakes in the first draft of this work.  This project arose from discussions during the KITP workshop on ``Bits, Branes, and Black Holes".  During this time it was therefore supported in part by the National Science Foundation under Grant No. NSF PHY11-25915. Much of the work on this paper was done at the 2014 Aspen Center for Physics workshops ``New Perspectives in Thermalization" and  ``Emergent Spacetime in String Theory"; the ACP is supported by National Science Foundation grant PHY-1066293.  A.L. and C.A. are supported in part by DOE grant DE-SC0009987; C. Agon was also supported in part by the National Science Foundation via CAREER Grant No. PHY10-53842 awarded to M. Headrick.   V.B. was supported by DOE grant DE-FG02-05ER-41367.

\eject

%\centerline{\bf {\Large Appendices}}
\addcontentsline{toc}{section}{Appendices}
\section*{Appendices}

\vskip .5cm

\appendix

\section{Relation to and differences from Wilsonian renormalization}

Textbook treatments of Wilsonian renormalization explicitly disentangle IR and UV degrees of freedom via a change of variables.   This point of view is important for computing the low-energy spectrum and the S-matrix of asymptotic states with low energies and long wavelengths.      Below we contrast this approach with the work in this paper.

%In standard treatments of Wilsonian renormalization, entanglement between the UV and the IR does not emerge as an issue.  These treatments give very accurate descriptions of the physical phenomena they are designed to describe.  Thus, the reader may be forgiven for thinking that UV-IR entanglement is irrelevant in discussions of the dynamics of coarse-grained degrees of freedom.  

%In fact, these treatments use a set of ``disentangled" variables, which are the physically relevant ones if one is trying to compute the low-lying energy spectrum, or compute S-matrix elements for the scattering of asymptotic particle states.  These are different questions from the ones we address in this paper. 

%In this appendix we elaborate on this point, discussing textbook versions of Wilsonian renormalization from various points of view, to draw a distinction from the work in the bulk of this paper. We also discuss the classical Born-Oppenheimer approximation as it relates to renormalization.  This is closest to the approach in our paper; in particular, once corrections to the Born-Oppenheimer approximation are included, the dynamics of the reduced density matrix ceases to be Hamiltonian.

\subsection{Path integral approach}

The standard discussion of Wilsonian renormalization ({\rm cf.} \cite{Wilson:1993dy,Peskin:1995ev}) begins with a Euclidean path integral
\be
	Z = \int D\phi(x) e^{-S_{\Lambda_0}[\phi]}
\ee
for a field theory with UV cutoff $\Lambda_0$.  One breaks up $\phi(x)$ into $\phi = \phi_s + \phi_f$, where the ``slow" fields $\phi_s$ are supported on Euclidean momenta $|k| < \Lambda$ and the ``fast" fields $\phi_f$ are supported on Euclidean momenta $\Lambda < |k|< \Lambda_0$. We coarse-grain the theory by integrating over $\phi_f$ to find
\be
	Z = \int D\phi_s e^{-S_{\Lambda}[\phi_s]}
\ee
For long-wavelength questions, we can work with this latter presentation.

When $Z$ represents the partition function in a classical equilibrium statistical physics problem,
the interpretation is clear: $S_{\Lambda}$ will represent the spatially coarse-grained classical Hamiltonian of the system. 
If we want to compute equal-time correlators in the analytical continuation to real-time, we need to impose periodic boundary conditions in the integral over the high frequencies,  while fixing the IR modes on two sides of a cut in time \cite{Balasubramanian:2011wt}.    This yields an effective action and an associated density matrix that can be used to compute equal time correlation functions.   But when $Z$ represents the quantum-mechanical vacuum-vacuum transition amplitude computed via Euclidean continuation, the decimation procedure above fixes both the initial and final state of the short-wavelength degrees of freedom. For the inclusive finite-time probabilities discussed in the introduction, this procedure is not appropriate, beyond the leading order in a Born-Oppenheimer approximation.  Note that when describing scattering of initially well-separated particle states into final states of the same form, the interactions are effectively inoperative at early and late times, and the assumption that the short-distance modes are in their ground state is essentially correct.   This standard treatment is designed to produce transition amplitudes between initial and final asymptotic states where only the low energy modes are excited.  By contrast, we are interested in finite-time questions for states that have UV-IR entanglement, including the natural ground states of interacting theories, and states produced by the action of coarse-grained operators.

For the inclusive finite-time questions we are discussing in this paper, the decimation procedure is best applied to spatial momenta, in the real-time path integral developed by Feynman and Vernon \cite{Feynman:1963fq}\ for density matrices.  In this case, the analog of the Wilsonian effective action will include terms describable as a renormalized Hamiltonian, together with a nontrivial ``influence functional" which encodes the time development of entanglement between the IR and UV degrees of freedom. We discuss this in Appendix \S{D}.

A further issue arises from the fact that higher-derivative interactions are generically induced.  These will include terms that are functions of ${\ddot \phi}_s$ and higher derivatives still, arising from nonlocalities on the scale of the running cutoff.\footnote{A related discussion, which partially inspired this paper, can be found in \cite{Balasubramanian:2011wt}.}  In general, the Wilsonian action $S_{\Lambda}$ will not have an interpretation as an action that can be derived from the Legendre transform of a Hamiltonian, unless one adds St\"uckelberg fields. Such a procedure amounts to adding the short-distance degrees of freedom back in. The interpretation of these higher-derivative terms is clear in the holographic picture of Wilsonian renormalization  \cite{Balasubramanian:2012hb}: they reflect the fact that the IR degrees of freedom comprise an open quantum system, and that there are memory effects on the time scale of the UV dynamics.

\subsection{Hamiltonian approach}

There is an alternative literature on Hamiltonian approaches to renormalization, pioneered originally by   Wilson \cite{Wilson:1965zzb,Wilson:1974mb}, and applied first to a model of pion-nucleon scattering and later to the Kondo problem, implemented by a successive diagonalization of degrees of freedom with a hierarchy of energy scales.  In these models the degrees of freedom of some quantum field are coupled through a localized defect. At each step one diagonalizes the Hamiltonian of the high energy degrees of freedom coupled to the defect, and works in the ground states of these degrees of freedom.  This diagonalization mixes the (iso)spin states of the defect with excitations of the high-energy modes of the quantum field: at each step, the low-energy spin degree of freedom becomes more delocalized.

Variants for interacting quantum fields (without a defect) can be found in, for example, \cite{Glazek:1993rc, Alexanian:1998wu}.  In \cite{Glazek:1993rc} one removes divergences by making a transformation to ``band-diagonal" form in which the Hamiltonian has no matrix elements between states with an energy difference larger than some value.  In \cite{Alexanian:1998wu}\ one implements partial diagonalization of the Hamiltonian by removing only the matrix elements between the IR band and the high energy degrees of freedom.

As in Wilson's work, the goal of the Hamiltonian approach to renormalization is to extract the spectrum.  To do so, we reorganize the theory in terms of effective low-energy degrees of freedom where the original low-frequency components of the Hilbert space are appropriately dressed by the high-frequency components so as to partially diagonalize the Hamiltonian between the UV and the IR.   
 If one is studying thermodynamics at low temperatures, or the dynamics of quasiparticles built from the renormalized variables, the low-energy Hilbert space can be treated as a closed quantum system.  Similarly, such an approach is also appropriate for S-matrix elements of well-separated particles.  
In this case, the initial state lies in the low-energy Hilbert space, and the final state will as well.  In this way, if one studies the scattering into final states of well-separated particles (or low-energy bound states), one can work entirely within the closed quantum system of the low-energy Hilbert space.

The calculations we described in this paper, however, assume that  measuring devices  couple to the bare variables with some finite spatial resolution, and that measurements are made at finite time.  In this setting, low- and high-momentum modes cannot be easily separated, and thus the measurable degrees of freedom form an open quantum system.

\subsection{Born-Oppenheimer approximation}

Our treatment of long-wavelength modes is closest to the Born-Oppenheimer approximation.  In textbook form \cite{messiah1999quantum,Moody:1989vh}, one separates the quantum mechanical degrees of freedom into ``fast variables" $Y$ and ``slow variables" $x$.  To implement the approximation, one considers the case that eigenstates of $x$ form a (possibly overcomplete) basis of the Hilbert space described by the slow degrees of freedom, and considers Hamiltonians of the form
\be
	H = H_x + H_Y(x)
\ee
For example, $x$ could be the positions of heavy nuclei, and $Y$ the positions of electrons moving in the backgrounds of these nuclei. 

% Consider $H_Y(x)$ for $x$ a $c$-number. 
Consider $x$ to take some frozen value and treat it as a background field.
 Then the Hilbert space of the ``fast" degrees of freedom can be written in eigenstates $\ket{n,x}$
\be
	H_Y(x) \ket{n,x} = E_n(x)\ket{n,x}
\ee
Let $\Delta E_x$, $\Delta E_Y(x)$ be the gap between eigenvalues of $H_x$, $H_Y(x)$. The simplest version of the Born-Oppenheimer approximation works when $\Delta E_Y \sim E_n(x) - E_m(x) \gg \Delta E_x$, for all $x$ where the wavefunction of the slow degrees of freedom has appreciable support.
Let $E_0(x)$ be the instantaneous ground state.  One can write the general wavefunction as:
%
%\be
%	\ket{\Psi(t)} = \int dx \psi(x,t) \ket{x}_x\ket{0,x}_Y + \sum_{n > 0} \int dx \delta\psi_n(x,t) \ket{x}_x \ket{n,x}_Y\ .
%\ee
\be
	\ket{\Psi(t)} = \int dx' \,  \psi(x',t) \,  \ket{x'}_x \ket{0;x'}_Y + \sum_{n > 0} \int dx' \, \delta\psi_n(x,t) \, \ket{x'}_x \ket{n;x'}_Y\ .
\ee
where the subscripts $x$ and $Y$ on the kets indicate states in the``slow" and ``fast" Hilbert spaces labeled by the indicated quantum numbers, while $\psi$ and $\delta\psi_n$ are the weights of the linear combination defining the full state.   A state like $|n,x'\rangle_Y$ indicates that the ``fast" modes $Y$ are in an energy eigenstate of $H_Y(x')$ with quantum number $n$.   This ``fast" eigenstate depends on the``frozen" value $x'$ of the slow variable through the dependence in $H_Y(x')$.  Meanwhile $\ket{x'}_x$ indicates a state in the ``slow'' Hilbert space  indexed by the slowly changing value $x'$.

To lowest order in the Born-Oppenheimer approximation $\Delta E_x/\Delta E_Y \ll 1$, the leading $n = 0$ term satisfies the time-dependent Schr\"odinger equation with effective Hamiltonian
\be
	H_{eff} = H_x + H_{Berry} + E_0(x)\label{eq:boheff}
\ee
where $H_{Berry}$ are the additional terms induced by Berry's phase \cite{Berry:1984jv,Simon:1983mh,Moody:1989vh}.\footnote{In the case that there are $N$ near-degenerate eigenstates with energies close to $E_0$, $\psi_0$ is replaced by an $N$-component wavefunction, with a non-Abelian $U(N)$ Berry's phase \cite{Wilczek:1984dh}.}  In this approximation, the reduced density matrix for the slow degrees of freedom can be written as
\be
	\rho_{IR} = \int dx' dx'' \psi_0(x')\psi_0^*(x'')  \,  \tr_Y \left[  \,\, \ket{x'}_x \,  \ket{0;x'}_Y \, {}_{Y}\bra{0,x''} \, {}_{x}\bra{x''} \, \right] 	\, .
%	\ket{x}\bra{y} \tr_Y \ket{x,0}\bra{y,0}
\ee
This describes a mixed state if ${\cal F}(x',x'') = \tr_Y \left[ \ket{x',0}_{Y}{}_{Y}\bra{x'',0} \right]$ is not factorizable in $x'$ and $x''$. Nonetheless, its evolution is unitary, with Hamiltonian $H_{eff}$, in this approximation.  The failure of unitarity -- that is, the status of the IR degrees of freedom as an {\it open}\ quantum system -- will appear at higher orders in the Born-Oppenheimer approximation, for finite-time processes.  This includes processes like recoil of the heavy degrees of freedom.   This has been discussed in the classical limit of the IR degrees of freedom in \cite{2014AnPhy.345..141D,2014arXiv1405.2077D}; corrections to the leading adiabatic limit lead to friction and dissipation.

This framework is essentially what we desire.  However, we are interested in the more general case of systems for which the coupling between IR and UV degrees of freedom cannot be simply expressed in terms of an IR operator which can be diagonalized. An example of this is the Hamiltonian for two coupled spins,
\be	
	H = - \mu_{L} B S_L^z - \mu_H B S_H^z + \lambda {\vec S}_L \cdot {\vec S}_H
\ee
where $\mu_L \ll \mu_H$, and the total spin ${\vec S}_L^2/\hbar^2  = j_L (j_L + 1)$ is not too large.  There is no basis which diagonalizes ${\vec S}_L$. On the other hand, for $j_L \gg 1$, or for long-wavelength modes in a spin chain, there is a semiclassical limit in which the spin can be treated as a semiclassical variable.  

\section{Entangled initial states}

When the initial state is entangled between the UV and IR, evolution of the IR density matrix is harder to characterize.  (See sec. 4 of \cite{2014RPPh...77i4001R}\ for a preliminary discussion of this case.) Let us consider the specific initial state
\be
	\ket{\Psi(0)} = \frac{1}{\sqrt{2}} \left( \ket{\chi} \ket{u_1} + \ket{\zeta} \ket{u_2} \right)
\ee
where $\ket{u_{1,2}}$ are eigenstates of $H_{UV}$ with eigenvalues $E_{1,2}$, and $\ket{\chi}, \ket{\zeta}$ are states in $\HIR$ which we will take to be linearly independent. The initial density matrix is
\be
	\rho(0) = \half \left( \ket{\chi}\bra{\chi} + \ket{\zeta}\bra{\zeta} \right)
\ee
As we will see, the complication will arise because each term will evolve differently, in a fashion dependent on the UV eigenstates they are coupled to.	
At zeroth order in $\lambda$, the density matrix is simply
\begin{eqnarray}
	\rho^{(0)}(t) &  = & \half e^{-i H_{IR} t} \left(\ket{\chi}\bra{\chi} + \ket{\zeta}\bra{\zeta} \right)e^{i H_{IR} t}\nonumber\\
	& = & \half \left( \ket{\chi(t)_I}\bra{\chi(t)_I} + \ket{\zeta(t)_I}\bra{\zeta(t)_I}\right)
\end{eqnarray}
where $\ket{\psi(t)_I} = e^{- i H_{IR} t} \ket{\psi(0)}$.  $\rho^{(0)}(t)$ evolves by Hamiltonian evolution, $i\hbar \p_t \rho^{(0)}(t) = [H_{IR}\,,\rho^{(0)}(t)]$.

At first order in $\lambda$, a calculation identical to those of section \ref{2} yields:
\begin{eqnarray}
	i \hbar \p_t \rho^{(1)}(t) & = & [H_{IR}, \rho^{(1)}(t)] \nonumber\\
	& & \ \ \ \ + [V_{11}, \ket{\chi(t)_I}\bra{\chi(t)_I} + [V_{22}, \ket{\zeta(t)_I}\bra{\zeta(t)_I}] \nonumber\\
	& & \ \ \ \ + [V_{12}e^{-i E_{21}t}, \ket{\zeta(t)_I}\bra{\chi(t)_I} + [V_{21} e^{i E_{21} t}, \ket{\chi(t)_I}\bra{\zeta(t)_I}] \nonumber \\
	\label{eq:nonfactmaster}
\end{eqnarray}
where $V_{ij} = \bra{u_i} V \ket{u_j}$, and $E_{21} = E_2 - E_1 \sim E_{UV}$.
There is no obvious sense in which the evolution is Markovian.  

The time averaging of the first-order evolution equation (\ref{eq:nonfactmaster}) is straightforward -- we simply drop the final line, which oscillates rapidly at a time scale of order $1/E_{UV}$.  The resulting equation is:
\be
	i\hbar \p_t{\overline \rho}(t) = [H_{eff,1}, \half \overline{\ket{\chi(t)}\bra{\chi(t)}}] + [H_{eff,2}, \half \overline{\ket{\zeta(t)}\bra{\zeta(t)}}]
\ee
where
\be
	H_{eff,i} = H_{IR} + \bra{u_i} V \ket{u_i}
\ee
This is not a Hamiltonian evolution.  

\section{UV-IR entanglement}

Having computed the density matrix, we can ask how entangled the systems become with time.  The most robust quantity to compute is the von Neumann entropy 
\be
	S(t) = - \Tr \rho(t) \ln \rho(t)
\ee
This can be difficult to compute in practice.  A simpler set of quantities to calculate are the R\'enyi entropies for $\rho(t)$:
\be
	S_{n}(t)=-\frac{\ln \textrm{Tr}\rho^n(t)}{n-1}\,.
\ee
If the resulting expression yields a smooth $n \to 1$ limit, one may use these to compute the von Neumann entropy.

We must take some care computing $S$ in perturbation theory, due to the logarithm.  If the unperturbed density matrix has zero eigenvalues and the perturbation is sufficiently generic, we expect the full density matrix to have eigenvalues that scale as $\lambda^p$.   Thus, there will be terms that scale as $\lambda^p \ln \lambda$ in the von Neumann entropy, and perturbation theory will break down: this fact was discussed in \cite{Balasubramanian:2011wt}.  While the R\'enyi entropies for fixed integer $n > 1$ can have good analytic expansions in $\lambda$, it is straightforward to see that the $\lambda \to 0$, $n \to 1$ limits will not commute. For a simple example, consider the density matrix
\be
	\rho = \left( \begin{array}{ll} 1 - a\lambda & 0 \\ 0 & a\lambda \end{array} \right)
\ee
For which
\begin{eqnarray}
	S_n & = & - \frac{1}{n-1} \ln \left[ \left(1 - a\lambda)^n + (a \lambda)^n\right) \right] \nonumber\\
	& = &  \frac{1}{n-1} \ln \left[ \left(1 - a\lambda) e^{(n-1) \ln (1 - a\lambda)} + (a \lambda) e^{(n-1) \ln(a\lambda)} \right) \right]
\end{eqnarray}

Note that in the cases we are studying, these entropies capture both the degree to which the initial IR density matrix is in a mixed state, as well as any entanglement that arises from time evolution of the coupled system. 
 Therefore, the most interesting question for us is the evolution of these quantities with time. Focusing on the R\'enyi entropies with integer $n$ (so that we are sure to work with well-defined quantities), we find:
\be
	\frac{dS_n(t)}{dt}=\frac{i n}{n-1}\frac{\textrm{Tr}[\rho^{n-1}(t)i\partial_t \rho(t)]}{\textrm{Tr}\rho^n(t)}\
\ee
%
%Note that in the limit $n \to 0$, the trace becomes $i \Tr \p_t \rho = i \p_t \Tr \rho = 0$ by conservation of probability, so the limit $n \to 1$ is potentially well defined. 
 If we insert (\ref{eq:tlocevol}), the contributions from $H_{eff}$ will vanish, due to the cyclicity of the trace, so that:
\be
	\frac{dS_n(t)}{dt}=\frac{in}{n-1}\frac{\textrm{Tr}[\rho^{n-1}(t)\Gamma(t)]}{\textrm{Tr}\rho^n(t)}\ \label{eq:secondorderrenyi}
\ee
Thus we see that the non-Hamiltonian components of the time-evolution specified by $\Gamma$ in (\ref{eq:tlocevol}) are precisely responsible for producing UV-IR entanglement as time passes.

Let us focus on the particular case that the initial IR state is an energy eigenstate $\ket{i}$ of the IR Hamiltonian, and work to $\CO(\lambda^2)$.  Since $\Gamma$ is nonvanishing only at $\CO(\lambda^2)$, we can evolve $\rho(t)$ with $H_{IR}$ alone, and it will remain pure.  Therefore we can replace $\rho^{n - 1} \to \rho$ for $n > 1$.  Then, using our known expressions for $\Gamma$, we find:
\begin{eqnarray}
	\frac{dS_n(t)}{dt} & = & \frac{2n}{n-1}\sum_{u\neq \bu ,\bj\neq \bi}\frac{\sin{\omega_{\bu u, \bi \bj}t}}{\omega_{\bu u, \bi \bj}}|\langle \bu,\bi|V|u,\bj\rangle|^2\nonumber\\
	S(t) & = & \int_0^t dt' \frac{dS_n(t')}{dt}\nonumber\\
&=& \frac{2n}{n-1}\sum_{u\neq \bu ,\bj\neq \bi}\frac{1-\cos{\omega_{\bu u, \bi \bj}t}}{\omega^2_{\bu u, \bi \bj}}|\langle \bu,\bi|V|u,\bj\rangle|^2\nonumber\\
	{\overline S}(t) & = & \frac{2n}{n-1}\sum_{u\neq \bu ,\bj\neq \bi}\frac{1}{\omega^2_{\bu u, \bi \bj}}|\langle \bu,\bi|V|u,\bj\rangle|^2
\end{eqnarray}
The R\'enyi entropies thus vary on the time scale of the UV degrees of freedom.  Note that $S_n(t) \geq 0$ always: thus the time average is nonvanishing and also time-independent (because the oscillations are at the UV timescale and the IR state is an eigenstate of the unperturbed Hamiltonian). 

\section{Path integral formalism}

The  time evolution of the reduced density matrix has a path integral formalism going back to Feynman and Vernon \cite{Feynman:1963fq,feynman2012quantum}, which points to an avenue for a systematic computation of higher-order corrections. Related results appear in the literature (see \cite{breuer2007theory}\ for a discussion and references).

Many readers may be familiar with this formalism in the context of quantum Brownian motion \cite{Caldeira:1982iu}\ and quantum dissipation \cite{Caldeira:1981rx,Caldeira:1982uj,Leggett:1987zz}; the coupled oscillator model (\ref{eq:coupledoscmodel})\ is a classic example to which this formalism has been applied.  As we discussed in \S3.4.2, the ``bath" of oscillators $x_{\ell}$ in these models contains a continuum of oscillators down to low frequencies, with a spectrum designed so that energy is dissipated into the bath without returning to the observed system over the lifetime of the experiment. Furthermore the bath is typically taken to be at finite temperature.  In our discussion, the ``bath" consists of degrees of freedom with high frequencies, which are generally in the ground state at leading order in $\lambda$.

\subsection{Review of the influence functional}

We wish to compute the density matrix starting with some known state $\ket{0}$ of the full system that evolves  forward in time.  The density matrix should express the probability that the final IR state is $\ket{\psi}$.  This can be written as:
\be
	P_{\psi} = \sum_u \bra{u}\bra{\psi} U(t,0) \ket{0}\bra{0} U(t,0)^{\dagger} \ket{\psi} \ket{u}
\ee
The amplitude $\bra{u}\bra{\psi} U(t,0) \ket{0}$ can be expressed as a path integral with the boundary conditions at times $0,t$ integrated against the wavefunction for $\ket{0}, \bra{u}\bra{\psi}$.  Similarly, the amplitude $\bra{0} U(t,0)^{\dagger} \ket{\psi} \ket{u}$ would be represented as the complex conjugate of this path integral: when the system enjoys time reversal invariance, this can be described in terms of paths propagating backwards in time.  The result is a path integral over paths moving forward then backwards in time, with the UV degrees of freedom at $t$ set equal and summed over.  This is reminiscent of the Schwinger-Keldysh formalism for ``in-in" expectation values \cite{Schwinger:1960qe,Bakshi:1962dv,Bakshi:1963bn,Keldysh:1964ud}: indeed, if one was to sum the above expression over all final states, one would arrive at the path integral expression for such expectation values.

To make this discussion more explicit, consider a system for which $\HIR$ describes the states of a particle with position $X$, and $\HUV$ the states of a particle with position $x$.  We consider an action of the form:
\be
	S[x,X]=S_U[{\dot x}, x]+S_I[{\dot X}, X]+\delta S[x,X]
\ee
where we write the interaction $\delta S$ in the form
\be
	\delta S[x,X]= - \lambda \sum_a\int_0^{t} dt' {\cal{O}}^{UV}_a\Phi^{IR}_{a}= - \lambda \sum_a\int_0^{t} dt' {\cal{O}}_a[x]\Phi_{a}[X] \label{eq:opexpand}
\ee
We are assuming the interaction term depends on the coordinates only and not on the velocities; thus the correction to the action is minus the correction to the Hamiltonian. We have factored out a small dimensionless parameter $\lambda \ll 1$ to better organize a perturbative treatment of the system.
As in the previous section, we also assume that the initial state of the system can be described by a factorized density matrix
\be
\sigma(x,X; y,Y; t = 0) = \rho_{IR}(X,Y; 0) \rho_{UV}(x,y; 0)
\ee

The density matrix for the IR degrees of freedom $X$ is:
\begin{eqnarray}
\rho(X,Y,t)&=&\int dX' dY' J(X,Y,t;X'Y',0)\rho_{IR}(X',Y',0)\,, \nonumber\\
J(X,Y,t;X'Y',0)&=&\int_{Y(0)=Y';\,X(0)=X'}^{Y(t)=Y;\,X(t)=X}{\cal D} X {\cal D} Y e^{\frac i{\hbar}S_I[X]-\frac i{\hbar}S_I[Y]}{\cal F}[X(t),Y(t)]\,,\nonumber\\ \label{eq:fpidensity}
\end{eqnarray}
where $\rho_{IR}$ is the initial density matrix for the IR degrees of freedom, and the {\it influence functional} 
\begin{eqnarray}
	\CF[X(t),Y(t)]&=&\int dx' dy' dx \rho_{UV}(x',y',0) \nonumber \\ 
	&\times &\int_{y(0)=y';\,x(0)=x'}^{y(t)=x;\,x(t)=x} {\cal D}x{\cal D}y e^{\frac i{\hbar}S_U[x]-\frac i{\hbar}S_U[y]+\frac i{\hbar}\delta S[x,X]-\frac i{\hbar}\delta S[y,Y]}\nonumber\\ \label{eq:influencedef}
\end{eqnarray}
contains the dependence on the initial state of the UV degrees of freedom, as well as the interactions between the UV and IR degrees of freedom. We will now pass to computing $\CF$ to order $\CO(\lambda^2)$.

\subsection{Perturbation theory for the influence functional}

We expand (\ref{eq:influencedef}) to second order in $\delta S[x,X] - \delta S[y,Y]$ to find 
\be	
	\CF[X,Y] = \CF^{(0)} + \lambda \CF^{(1)}[X,Y]  + \lambda^2 \CF^{(2)}[X,Y]
\ee
using the representation (\ref{eq:opexpand}). Up to first order, we find
\begin{eqnarray}
	\CF^{(0)} & = & 1\nonumber\\
	\CF^{(1)} & = & \frac{ - i}{\hbar}  \sum_a \int_0^t dt' \vev{{\cal O}_a^I(t')} \left( \Phi_{a}[X]-\Phi_{a}[Y]\right)\nonumber
\end{eqnarray}
where 
\be
	\vev{{\cal O}_a^I(t')} \equiv \Tr_{_{UV}} \left(  {\cal O}_a^I(t') \rho_{UV}(0)\right) = \Tr_{UV} \left({\cal O}_a^I(t' - t) \rho_{UV}(t)\right) \equiv \vev{{\cal O}_a^I(t'-t)}_t
\ee
The superscript $I$ denotes the interaction picture, and $\rho_U(t)$ evolves as 
\be	
	i \p_t \rho_{UV}(t) = [H_{UV},\rho_{UV}(t)]
\ee
in this expression. 
%${\cal F}^{(1)}$ can be absorbed into a first-order correction to the IR action:
%%
%\be
%	S^{(1)}_{IR} = - \lambda \int_0^t dt' \vev{{\cal O}_a^I(t')}_0 \Phi_a(t')
%\ee
%%
%which is equivalent to the first-order correction (\ref{eq:firstorderham}).
%

The second order correction is:
\begin{eqnarray}
	{\cal F}^{(2)}&=& - \frac{\lambda^2}{2\hbar^2} \sum_{a,b} \int_0^t dt' dt'' P^F_{a,b}(t',t'')\Phi^I_{a}[X(t')]\Phi^I_{b}[X(t'')]\nonumber   \\
 & & - \frac{\lambda^2}{2\hbar^2} \sum_{a,b}\int_0^t dt' dt'' \widetilde{P}^F_{a,b}(t',t'')\Phi^I_{a}[Y(t')]\Phi^I_{b}[Y(t'')]\nonumber   \\
 & & + \frac{\lambda^2}{\hbar^2} \sum_{a,b} \int_0^t dt' dt'' P^W_{a,b}(t',t'')\Phi^I_{a}[X(t')]\Phi^I_{b}[Y(t'')]\,.\label{eq:secondinfluence}
 \end{eqnarray}
 where
 \begin{eqnarray}
 	P^F_{a,b}(t',t'')&=&\Tr_{_{UV}} T({\cal O}^I_a(t'){\cal O}^I_b(t''))\rho_{UV}(0)\,, \\
	\widetilde{P}^F_{a,b}(t',t'')&=&\Tr_{_{UV}} \widetilde{T}({\cal O}^I_a(t'){\cal O}^I_b(t''))\rho_{UV}(0)\,, \\
	P^W_{a,b}(t',t'')&=&\Tr_{_{UV}} {\cal O}^I_a(t'){\cal O}^I_b(t'')\rho_{UV}(0)\,,
\end{eqnarray}
and $\widetilde{T}$ denotes time anti-ordering.  Note that the time-ordered two-point function, the anti-time-ordered two-point function, and the Wightman function appear for essentially the same reason that they do in the Schwinger-Keldysh formalism for in-in expectation values. Finally, note that all of the operators should be understood as being in the interaction picture.

At $\CO(\lambda^2)$, we can exponentiate the $\CO(\lambda)$ term, to arrive at:
\be
	\CF = e^{\lambda  \CF^{(1)}} \left(1 + \lambda^2 {\tilde \CF}^{(2)}\right) + \CO(\lambda^3)\label{eq:reexp}
\ee
The effect is to shift $P^F$, $\widetilde{P}^F$, and $P^W$ to $G^F$, $\widetilde{G}^F$, $G^W$, where
\be
	G^{F,W}_{a,b}(t,t') = P_{a,b}^{F,W}(t,t') - \vev{O_a(t)}\vev{O_b(t')}
\ee
and $\widetilde{G}^F = (G^F)^*$ as before. 

\subsection{Relating the influence functional to the master equation}

If we insert (\ref{eq:reexp}) into (\ref{eq:fpidensity}), $\CF^{(1)}$ can clearly be absorbed into a shift in the action of the form
\be
	\delta S^{(1)}[X] = - \int^t_0 dt'_0 \lambda \vev{O_a(t')} \Phi_a[X(t')]
\ee
As a shift in the Hamiltonian, this is identical to the result (\ref{eq:firstorderham}) derived via operator methods, if the initial state of the UV degrees of freedom is the pure state $\ket{\bu}$.

Next, by taking the time derivative of (\ref{eq:fpidensity}) using (\ref{eq:reexp}), we can write the master equation to this order in terms of UV Green functions:
\begin{eqnarray}
	i \p_t \rho^{(2)}(t) & = & [H_{IR}, \rho^{(2)}] + [H^{(1)}, \rho^{(1)}] \nonumber\\
	& & \ \ \ \ \ + i \lambda \int_0^t d\tau \, G^W_{ab}(\tau,t) \, \Phi_b(0) \rho^{(0)}(t)\, \Phi_a(\tau - t)\nonumber\\
	& & \ \ \ \ \ + i \lambda \int_0^t d\tau \, G^W_{ab}(t,\tau) \, \Phi_b(\tau-t) \rho^{(0)}(t) \Phi_a(0)\nonumber\\
	& & \ \ \ \ \ - i \lambda \int_0^t d\tau \, G^W_{ab}(t,\tau) \, \Phi_a(0) \Phi_b(\tau - t) \, \rho^{(0)}(t)\nonumber\\
	& & \ \ \ \ \ - i \lambda \int_0^t d\tau \, G^W_{ab}(\tau,t) \, \rho^{(0)}(t) \Phi_a(\tau-t) \, \Phi_b(0)
\end{eqnarray}
This can be written in the form (\ref{eq:tlocevol},\ref{eq:tlmaster})\ if we write $L_k = L_{am}$, $m\in\{1,2\}$, and set
\begin{eqnarray}
	h_{am;bm'} & = & \lambda^2 \delta_{ab}  |m- m'|
	%\left(\begin{array}{ll} 0 & \frac{1}{\tau^*} \\ \frac{1}{\tau} & 0 \end{array}\right)
	\nonumber\\
	L_{a1} & = &  \Phi_a(0) \nonumber\\
	L_{a2} & = & \int_0^t d\tau \, G^W_{ab}(t,\tau) \Phi_b(\tau-t)
\end{eqnarray}
(where again $\tau$ is a complex parameter which factors out of the master equation, but is included so that $L_{ai}$ all have the same dimension).\footnote{Note that if we were able to assume that $G^W_{ab}\propto\delta(\tau - t)$, these results would be consistent with Eq. (2.4-2.5) of \cite{Burgess:2014eoa}.}  Finally, the operators in (\ref{eq:tlocevol}) then become:
\begin{eqnarray}
	H^{(2)} & = & i \int^t_0 d\tau \left( G^W_{ab}(\tau,t) \Phi_a(\tau-t) \Phi_b(0) - G^W_{ab}(t,\tau) \Phi_a(0)\Phi_b(\tau-t)\right)\nonumber\\
	A^{(2)} & = & \int^t_0 d\tau \left( G^W_{ab}(\tau,t) \Phi_a(\tau-t) \Phi_b(0) + G^W_{ab}(t,\tau) \Phi_a(0)\Phi_b(\tau-t)\right)\nonumber\\
	\gamma^{(2)} & = & i \int^t_0 d\tau \left( G^W_{ab}(\tau,t) \Phi_b(0) \rho^{(0)}(t) \Phi_a(\tau-t) + 
		G^W_{ab}(t,\tau) \Phi_b(\tau-t)\rho^{(0)}(t) \Phi_a(0)\right)\nonumber\\
\end{eqnarray}
A calculation shows that these expressions are equivalent to those in Sec. 2.

\subsection{Criteria for Markovian behavior}

The expressions for $H^{(2)}$, $A^{(2)}$, and $\Gamma^{(2)}$ above indicate a necessary condition for time-local, Markovian evolution, namely that $G^W_{ab}$ (the unordered Wightman function) falls off rapidly for $|\tau - t| \geq \delta t$.  In typical quantum systems, this requires (see for example \cite{Barbon:2003aq,Dyson:2002pf}):
\begin{itemize}
\item The operators ${\cal O}$ should have matrix elements between the initial UV state ${\ket u}$ and a set of UV energy levels that are finely spaced by the inverse of a time scale $t_P$ much larger than the scale of the experiment; at scales of order $t_P$ one expects quasiperiodic behavior characteristic of Poincar\'e recurrences.
\item The matrix elements of ${\cal O}$ contributing to the correlation function should have a finite width $\Gamma$ in energy, leading to exponential falloff at a time scale $\Gamma^{-1}$. For finite-temperature correlators, where the Boltzmann factor cuts off large-energy states, this may be of order the inverse temperature.
\end{itemize}

Local correlators alone are not sufficient to guarantee behavior that is Markovian in the strict sense of the dynamical map being divisible.  As an example, for a Brownian particle coupled to a spectrum of harmonic oscillators at finite temperature $T_B$  \cite{Caldeira:1982iu}, correlators fall off on a time scale of order $T_B^{-1}$.  However, even on time scales long compared to $T_B^{-1}$, the master equation fails to be Markovian up to a term scaling as $\gamma/T_B$ (see \S3.6.2 of \cite{breuer2007theory}), where $\gamma$ controls the spectral density of the oscillators and sets the time scale for relaxation of the IR system.  For zero-temperature dynamics there is even less reason for Markovian dynamics to emerge.
Since our model two-scale systems do not satisfy the above assumptions, we do not expect Markovian behavior.  

%We reiterate that standard treatments of Wilsonian renormalization are not only Markovian but Hamiltonian, because they work with renormalized variables designed so that the Hamiltonian is factorized between low and high energies. However, we can imagine an experiment in which our apparatus couples to the spatially averaged bare variables.  Furthermore, we were inspired by the case of holographic renormalization; if the radial cutoff has anything to do with a split between IR and UV degrees of freedom, it is clear from \cite{Balasubramanian:2012hb}\ that the IR degrees of freedom in this case are an open quantum system. 

\eject
\bibliographystyle{utphys}
\bibliography{rtrgrefs}

\end{document}